\documentclass{aa}  

\def\4u{4U1820-303}

\def\be{\begin{equation}} 
\def\ee{\end{equation}}

\def\ergscm2{erg s$^{-1}$ cm$^{-2}$}
\def\xspec{\texttt{XSPEC}}

\def\jaxspec{\texttt{jaxspec}}

\def\sbi{\texttt{sbi}}

\usepackage{graphicx}
\usepackage{txfonts}
\usepackage{hyperref}
\usepackage{comment}
\usepackage{subfigure}
\usepackage{lscape}             
\usepackage{placeins}     
\begin{document} 

   \title{Simulation-based inference with neural posterior estimation applied to X-ray spectral fitting}

   \subtitle{II - High-resolution spectroscopy with the X-ray Integral Field Unit}
    \titlerunning{Neural networks for X-ray spectral fitting}
   \author{Simon Dupourqué
          \inst{1}
          \and
          Didier Barret\inst{1}}

   \institute{Institut de Recherche en Astrophysique et Planétologie, 9 avenue du Colonel Roche, Toulouse, 31028, France\\
              \email{sdupourque@irap.omp.eu}
              }

   \date{Received XXXX; accepted YYYY}

 
  \abstract
   {X-ray spectral fitting in high-energy astrophysics can be reliably accelerated using Machine Learning. In particular, Simulation-based Inference (SBI) produces accurate posterior distributions in the Gaussian and Poisson regime for low-resolution spectra, much faster than other exact approaches such as Monte Carlo Markov Chains or Nested Sampling.} 
   {We now aim to highlight the capabilities of SBI for high resolution spectra, as what will be provided by the newAthena X-ray Integral Field Unit (X-IFU). The large number of channels encourages us to use compressed representations of the spectra, taking advantage of the likelihood-free inference aspect of SBI.}
   {Two compression schemes are explored, using either simple summary statistics, such as the counts in arbitrary bins or ratios between these bins, or automatically learning compressed representation using dense neural networks. We benchmark the efficiency of these approaches using simulated X-IFU spectra with various spectral models, including smooth comptonized spectra, relativistic reflexion models and plasma emission models.} 
   {We find that using simple and meaningful summary statistics is much more efficient than working directly with the full spectrum or automatically-learned summary statistics, and can derive posterior distributions comparable to those from exact computation using nested sampling. Multi-round inference converges quickly to the good solution. Amortized single round inference requires more simulations, hence longer training time, but can be used to infer model parameters from many observations afterwards. We show an application where it can be used to explore the sensitivity of an observation to a model, for which the parameters spreads around the targeted model. Information from the emission lines must be accounted for using dedicated summary statistics.}
   {Simulation based inference for X-ray spectral fitting is a robust technique that delivers well calibrated posteriors. Although in its early days of development, the method shows great promises for high resolution spectra, offering its potential for the scientific exploitation of the X-IFU. We now plan to apply it to the current era of high-resolution telescopes, and further challenge this approach with real data.}

   \maketitle
%

\section{Introduction}

X-ray spectral fitting is a cornerstone of high-energy astrophysics. Traditional maximum likelihood estimation approaches define a fit statistic that describes the distribution of the observed spectrum as a function of the model parameters. In general, X-ray spectra can be modelled as a counting process and inference maximises a Poisson likelihood (strictly equivalent to minimising a Cash statistic or \texttt{Cstat}, see \cite{Cash1979ApJ...228..939C}). However, direct optimisation is an inefficient process when exploring a multidimensional likelihood landscape, relies heavily on proper initialisation, and can be quite unstable given its initial parameter estimates. This can lead to suboptimal results where the inferred parameters are given by a local extremum of the statistic, thus misleading the physical interpretation (e.g. \cite{bonson_how_2016, choudhury_testing_2017, Barret_2019A&A...628A...5B}). The Bayesian approach provides tools that are more robust to perform this inference task in high-dimensional parameter spaces, such as Markov Chain Monte Carlo \citep[MCMC,][]{hastings_monte_1970} or Nested Sampling \citep[NS,][]{10.1214/06-BA127}. Both of these methods are more robust to local extrema, but require significant computation power. Existing software for X-ray spectral fitting often uses direct minimisation or MCMC\citep{Arnaud1996ASPC..101...17A, isis_package, spex_package, siemiginowska_sherpa_2024}, whereas NS implementation are available in tools such as \texttt{BXA} \citep{Buchner2014A&A...564A.125B}. 

Accelerating the inference of X-ray spectra while maintaining reliable results will be challenging for the next decade of high-energy astrophysics, as observatories with unprecedented spectral resolution such as XRISM/Resolve \citep{xrismscienceteam2020sciencexrayimagingspectroscopy} and \textit{newAthena}/X-IFU \citep{Barret2023ExA....55..373B, peille-xifu-20205} will produce high-quality spectra to be exploited. A first approach would be to use what we have learned from several decades of advances in machine learning (ML). In particular, using hardware acceleration to perform the computation on GPUs and writing a differentiable likelihood function can leverage the use of powerful methods such as variational inference or Hamiltonian Monte Carlo \citep{betancourt2013hamiltonianmontecarlohierarchical}. Such an approach is already under development in the \texttt{jaxspec} package \citep{Dupourque2024A&A...690A.317D}. A second approach would be to use ML directly, and in particular Neural Networks (NN), to perform inference using approximate representations. This has been explored by \cite{Ichinohe2018MNRAS.475.4739I} and \cite{Parker2022MNRAS.514.4061P} respectively for the study of the Intra-Cluster Medium (ICM) and Active Galactic Nuclei (AGN) for respectively high and low-resolution spectra. They showed that NNs can constrain the parameter values with a lower risk of local minima compared to direct minimisation. Nowadays, NNs are used not only to estimate parameter values but also to directly characterize the associated uncertainties. \cite{tutone2025xrayspectralfittingmonte} successfully explored the use of Monte Carlo dropout to include randomness in the reconstruction and provide a distribution instead of single values. In \cite{Barret_2024AA} (hereafter Paper I) we used Simulation-Based Inference (SBI) with Neural Posterior Estimation (NPE) to estimate the posterior distribution of the parameters. This approach relies on normalising flows, which are parametrized transformations between probability distributions that can learn a mapping between the distribution of the observable and the distribution of the parameters. We demonstrated the technique to work equally well as standard X-ray spectral fitting, both in the Gaussian and Poisson regimes, on simulated and real data, yielding fully consistent results in terms of best-fit parameters and posterior distributions, with classical software fitting package such as \texttt{XSPEC} \citep{Arnaud1996ASPC..101...17A}. The inference time is comparable to or smaller than the one needed for Bayesian inference when involving the computation of large Markov chain Monte Carlo (MCMC) chains to derive the posterior distributions. On the other hand, once properly trained, an amortized SBI-NPE network generates the posterior distributions in less than a second. 

In this paper, we aim to push forward the application of SBI-NPE. First, we recall briefly the methodology to produce the simulated data in Sec.~\ref{methodology}. To demonstrate the power of likelihood free inference, beyond the principal component decomposition discussed in Paper I, we consider alternative ways of reducing the data dimensions, either through summary statistics or through an embedding network. We explore iterative approaches with multiple rounds of inference, and demonstrate how SBI-NPE is well suited for conducting feasibility studies with X-IFU thanks to single round inference that produces amortized networks. We then demonstrate the applicability of the technique to state-of-the-art models with emission lines. All the spectra displayed in this paper are simulated high-resolution spectra, as will be provided by the X-IFU spectrometer on-board \textit{newAthena} \citep{cruiseNewAthenaMissionConcept2025}.

\begin{table}
\caption{Parameters true value and prior distribution used in the three models described in this paper.}
\label{tab:model-parameters}
\begin{center}
\texttt{tbabs*(comptt+powerlaw)}
\begin{tabular}{llll}
\hline
Component & Parameter & Truth & Prior \\
\hline
\hline
\texttt{tbabs} & $n_H$ & 0.1 & $\mathcal{U}(0.01, 0.2)$ \\ 
\texttt{comptt} & $z$ & 0 & Frozen\\ 
\texttt{comptt} & $T_0$ & 0.55 & $\mathcal{U}(0.05, 2)$ \\ 
\texttt{comptt} & $kT$ & 2.5 & $\mathcal{U}(2.1, 4)$ \\ 
\texttt{comptt} & $\tau_p$ & 2.5 & $\mathcal{U}(0.5, 7)$\\
\texttt{comptt} & approx & 1 & Frozen\\ 
\texttt{comptt} & norm & 1 & $10^{\mathcal{U}(-1, 1)}$ \\ 
\texttt{powerlaw} & $\alpha$ & 2 & $\mathcal{U}(0, 5)$ \\ 
\texttt{powerlaw} & norm & 0.275 & $10^{\mathcal{U}(-1, 1)}$ \\ 
\hline
\end{tabular}

\vspace{5mm}
\texttt{tbabs*relxillNS}

\begin{tabular}{llll}
\hline
Component & Parameter & Truth & Prior \\
\hline
\hline
\texttt{tbabs} & $n_H$ & 0.2 & $\mathcal{U}(0.1, 0.3)$ \\ 
\texttt{relxillNS} & index$_1$ & 3 & Frozen\\ 
\texttt{relxillNS} & index$_2$ & 3 & Frozen\\ 
\texttt{relxillNS} & $R_{\text{br}}$ & 15 & Frozen\\ 
\texttt{relxillNS} & $a$ & 0.2  & $\mathcal{U}(0.05, 0.4)$ \\ 
\texttt{relxillNS} & $i$ & 30 & Frozen\\ 
\texttt{relxillNS} & $R_{\text{in}}$ & -1 & Frozen\\ 
\texttt{relxillNS} & $R_{\text{out}}$ & 400 & Frozen\\ 
\texttt{relxillNS} & $z$ & 0 & Frozen\\ 
\texttt{relxillNS} & $kT_{\text{BB}}$ & 1.05 & $\mathcal{U}(0.6, 2)$ \\ 
\texttt{relxillNS} & $\log \xi$ & 3.1 & $\mathcal{U}(1, 4.7)$ \\ 
\texttt{relxillNS} & $A_{\text{Fe}}$ & 2 & Frozen\\ 
\texttt{relxillNS} & $\log N$ & 16 & Frozen\\ 
\texttt{relxillNS} & $f$ & 2.5 & $\mathcal{U}(0.5, 4.5)$\\ 
\texttt{relxillNS} & norm & 0.01 & $10^{\mathcal{U}(-3, -1)}$ \\ 

\hline
\end{tabular}

\vspace{5mm}
\texttt{tbabs*(bapec+bapec)}

\begin{tabular}{llll}
\hline
Component & Parameter & Truth & Prior \\
\hline
\hline
\texttt{tbabs} & $n_H$ & 0.1 & Frozen \\
\texttt{bapec}$_1$ & $kT_1$ & 4  & $\mathcal{U}(3, 6)$ \\ 
\texttt{bapec}$_1$ & $Z_1$ & 1.2  & $\mathcal{U}(0.5, 2.5)$ \\ 
\texttt{bapec}$_1$ & $z_1$ & 0.02  & $\mathcal{U}(0, 0.05)$ \\ 
\texttt{bapec}$_1$ & $v_1$ & 120  & $\mathcal{U}(50, 200)$ \\ 
\texttt{bapec}$_1$ & norm$_1$ & 1  & $10^{\mathcal{U}(-0.3, 0.2)}$ \\ 
\texttt{bapec}$_2$ & $kT_2$ & 3  & $\mathcal{U}(1, 4)$ \\ 
\texttt{bapec}$_2$ & $Z_2$ & 1  & $\mathcal{U}(0.5, 2.5)$ \\ 
\texttt{bapec}$_2$ & $z_2$ & 0.02  & $\mathcal{U}(0, 0.05)$ \\ 
\texttt{bapec}$_2$ & $v_2$ & 100  & $\mathcal{U}(50, 200)$ \\ 
\texttt{bapec}$_2$ & norm$_2$ & 1.5  & $10^{\mathcal{U}(0, 0.3)}$ \\ 
\hline
\end{tabular}
\end{center}

\end{table}

\section{Methodology}
\label{methodology} 

Simulation‐based inference aims to train a neural network to learn an optimal approximation of the posterior distribution. Throughout this paper, we refer to the parameters as $\boldsymbol{\theta}$ and to the observation as $\boldsymbol{x}$. So, once trained, what we have is an effective approximation $q$ of the posterior distribution, i.e. $ q(\boldsymbol{\theta}) \simeq p(\boldsymbol{\theta} | \boldsymbol{x}) $. Unlike MCMC or NS, this approach does not require computing the likelihood function explicitly, and rather learn an approximation of it through a set of simulations. As in traditional inference approaches, it requires being able to compute the observed spectrum, folded by the instrument response, for a given set of parameters of the spectral model we want to fit. In Paper I, we performed these simulations using the \jaxspec~\citep{Dupourque2024A&A...690A.317D} software, which is a \texttt{JAX}-based X-ray spectroscopy software, optimized for the generation of many spectra at once on the GPU. However, \jaxspec~requires the model to be written in pure \texttt{JAX} to be used in the pipeline, which is not yet the case for many widely used spectral models such as the \texttt{relxill} or \texttt{APEC} family models. In this paper, we choose to use \xspec\ \citep{Arnaud1996ASPC..101...17A} instead, to take advantage of the largest set of available spectral models. Simulations are performed in parallel using the \texttt{PyXspec} wrapper and post-processed in \texttt{Python} to account for the Poisson noise. Traditional approaches would seek to minimise or explore the likelihood landscape as defined by the \texttt{Cstat} \citep[and should hold to this even for relatively high count spectra, see][]{Buchner2023arXiv230905705B}, assuming that the measured counts in each bin come from a single Poisson realisation of the expected number of counts. SBI approaches therefore require us to include this variance, which is achieved by simply performing a Poisson realisation of the predicted counts in each bin. This is crucial as it allows the network to learn the likelihood distribution, making the SBI approach asymptotically equivalent to MCMC/NS with \texttt{Cstat}.\footnote{Note that in this paper we have not covered the case where the spectra follow an effective Gaussian statistic, as provided by some reduction software. This could be achieved by simply switching from a Poisson realisation to a Gaussian realisation with the associated (co-)variance.}
In this article, we are interested in demonstrating the applicability of SBI for high-resolution X-ray spectroscopy, and in particular for the X-IFU instrument \citep{Barret2023ExA....55..373B, peille-xifu-20205}, with the updated response\footnote{\url{https://x-ifu.irap.omp.eu/en/science/instrument-response-files}} considering the 13 row mirror, a spectral resolution of 4 eV and the open filter position.

\begin{figure*}[t]
    \centering
    \includegraphics[width=0.49\linewidth]{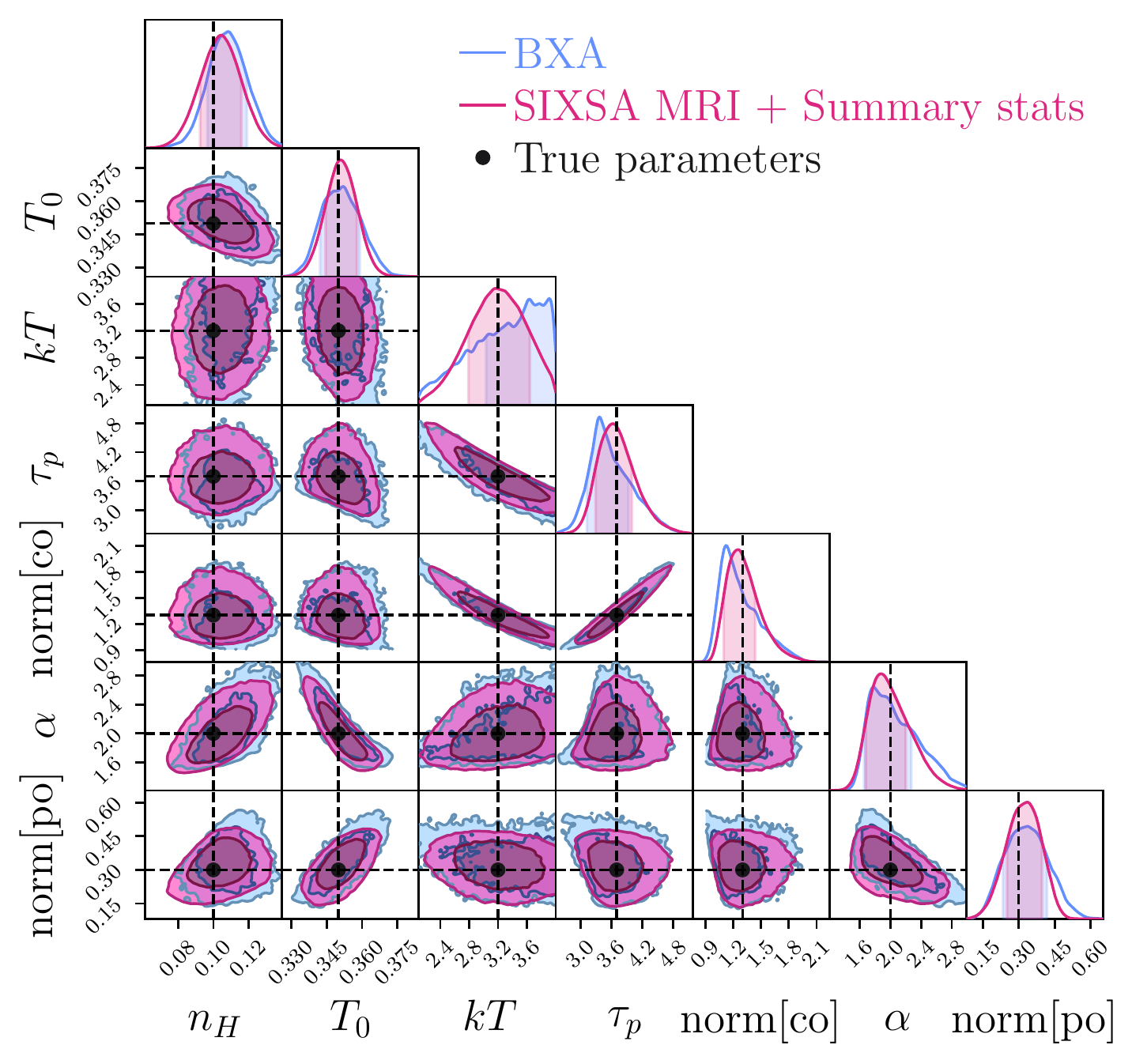}
    \includegraphics[width=0.49\linewidth]{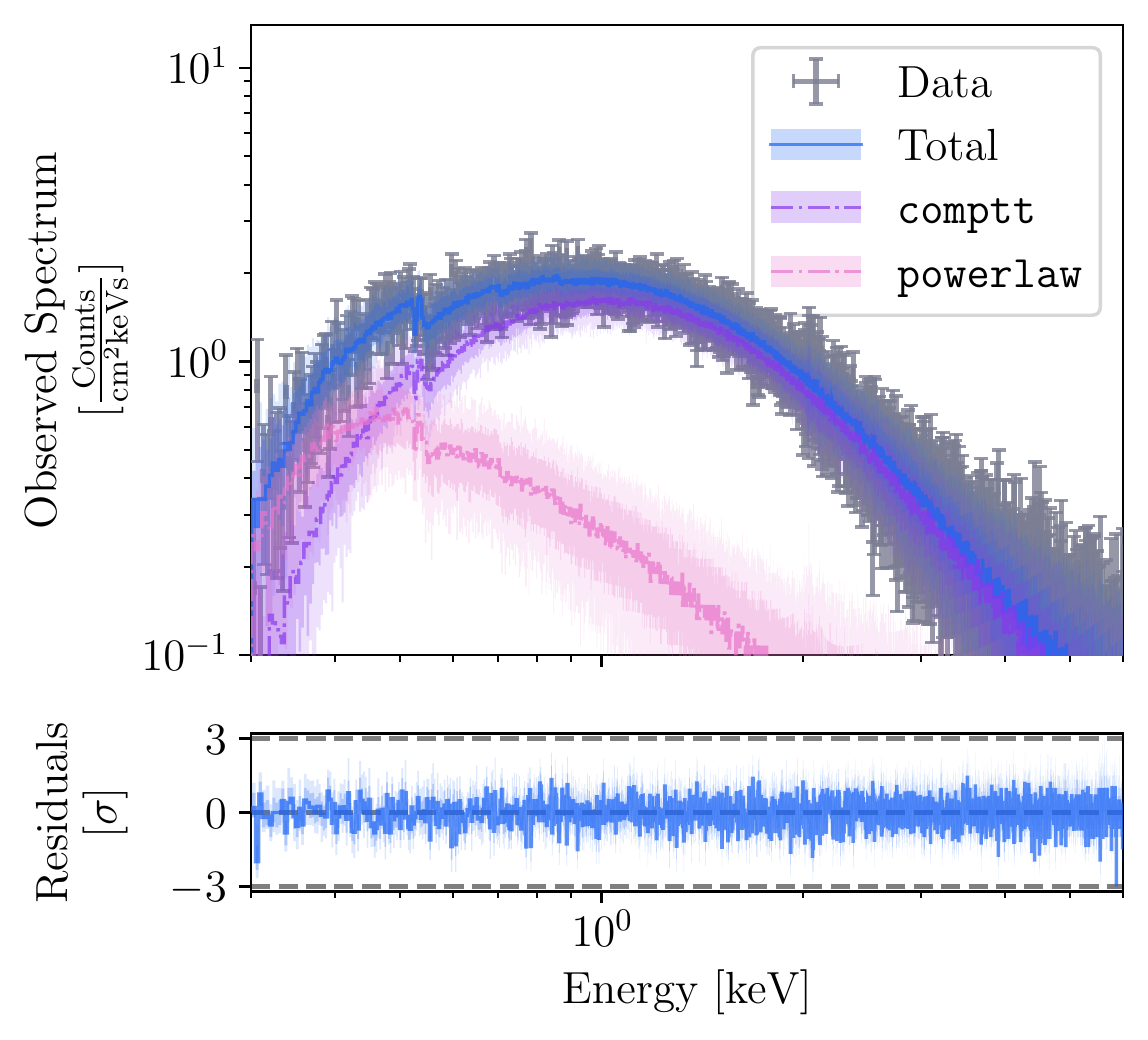}
    \caption{Left: Comparison of the posterior distributions obtained from a \texttt{BXA} run with 1k live-points and a MRI run with 5 rounds and 5k simulations each using summary statistics for a \texttt{tbabs*(comptt + powerlaw)} model. Right: Posterior predictive spectra associated with the posterior distribution obtained using summary statistics. The bands detail the (16-84) and (2.5-97.5) percentiles of the posterior predictive spectra.}
    \label{fig:mri-bxa}
\end{figure*}

In the same fashion as in Paper I, we rely on the \sbi\footnote{\url{http:://sbi-dev.github.io/sbi/}} \citep{Tejero2020JOSS....5.2505T} \texttt{Python} package. \sbi\ is a PyTorch-based package that provides a unified interface to state-of-the-art SBI algorithms together with very detailed documentation and tutorials. All the inferences presented in this paper uses Neural Posterior Estimation as defined by \cite{greenberg2019automaticposteriortransformationlikelihoodfree}, which can cope with proposal distributions that differ from the actual prior distribution of the inference problem. This is extremely useful when performing iterative SBI, where the parameter posterior distribution is refined after multiple rounds of simulations which get closer to the actual observation. Such approaches require probability density estimators, and the usual for this are neural networks. In particular, we use Masked Autoregressive Flow \citep[MAF,][]{papamakariosMaskedAutoregressiveFlow2017} as a density estimator with 10 transforms and 100 hidden units each. The size of the network has been increased from Paper I to reflect the increase in spectral resolution and data complexity. These hyperparameters are a good default set for all the inference problems treated in this paper. Increasing the number of transformations and the number of hidden parameters increases the overall expressiveness of the neural network, and enables it to learn more complex transformations. In addition, a low learning rate ensures high-quality training and avoids retaining sub-optimal network configurations. The training dataset is split in a 90/10 \% validation/test set. We consider the Adam optimizer \citep{kingma2017adam} with a $5\times10^{-4}$ learning rate, and patience of 20 epochs, and minimize the atomic loss defined by \cite{greenberg2019automaticposteriortransformationlikelihoodfree}. SBI comes in two flavours. One can perform Single-Round Inference (SRI) and train a single NN with numerous simulations to get an amortized NN that can be used with various spectra. Otherwise, Multi-Round Inference (MRI) will train a neural network with several rounds of inference and a reduced number of simulations. After each round, the simulations are drawn from the current surrogate posterior distribution, which shrinks toward the optimal values. When performing MRI, we use the truncated proposal distribution following \cite{deistler2022pnas2207632119}. Indeed, sampling from the iterative $\boldsymbol{\theta}$ distributions can become slow and inefficient when increasing the number of round. A truncated proposal trains an on-the-fly NN classifier that can instantly reject samples with a probability below a given threshold, dramatically improving sampling speed and efficiency. In each round, we train a truncated proposal to exclude samples outside the $10^{-4}$ quantile, which allows not only faster iterations, but also more inference rounds. All these methods are implemented in the simulation-based inference for X-ray Spectral Analysis (\texttt{SIXSA}\footnote{\url{https://github.com/dbxifu/SIXSA}}) software described in Paper I, including the use of summary statistics. The code used for this paper has been ported from the original codebase in a fork of \texttt{BXA}\footnote{\url{https://github.com/JohannesBuchner/BXA/}}, with an equivalent interface, and is available online\footnote{\url{https://github.com/renecotyfanboy/bsixsa}}. 

\section{Dimension reduction}

Simulation-based inference essentially finds a mapping between the parameter space and the observation space. In this respect, it suffers from the same problems as a lot of machine learning tasks, which have to find mappings between a low-dimensional representation (e.g. a few classes for classification) and the high-dimensional observation space. To overcome this, it is common to work with compressed and normalised representations of observables, which makes learning easier and increases the performance of neural networks, see e.g. \cite{ayesha_overview_2020} for a review. Since SBI is a likelihood-free approach, it can work with any meaningful transformation of the observable instead of the direct observable. In Paper I, we showed that reducing the dimensions of the spectra using principal component analysis is efficient. However, when working with high resolution spectral data, the number of components required to explain the variance of the data scales into the thousands. This parameter space is too large to map and leads to low performance when running SBI. This is also the case when using reduction schemes such as wavelet transform without proper coefficient selection or fine-tuning to the specifics of X-ray spectroscopy. In the following section, we would like to explore various compression schemes and assess the effectiveness of this approach in different work cases. 

\subsection{Summary statistics}
\label{sec:summary}
The first way of compressing data that we chose to investigate is probably as simple as it is powerful. The count information of an X-ray spectrum, spread across its bins, can be reduced to a set of statistical quantities such as the total number of counts, the mean number of counts, the standard deviation, the number of counts in adjacent energy intervals, the hardness ratios in adjacent energy bands, and so on. We refer to this set of quantities as the summary statistics for this inference problem. By reducing the spectrum using summary statistics that encode normalisation and overall shape, model training becomes much faster and more efficient. For an X-IFU spectra, we commonly go from 20k channels to only a few tens of features. This is mainly due to the summary statistics being less numerous and covariant than the original spectrum, and subsequently, it reduces the overall marginality of the spectrum we want to fit within the simulation set, allowing the network to interpolate better. To get a compressed and meaningful representation of the spectra, we use simple and tractable summary statistics. The simplest statistics we use are the mean, sum and standard deviation of the counts estimated on the whole spectrum, which carry information about the normalisation and noise of the spectrum. To this, we add quantities evaluated on a coarse energy grid. The spectrum is split in 50 logarithmic spaced bins, and we compute the sum of the counts in each of these bins. We also compute the hardness ratios HR and what we call the differential ratios DR between these bins defined by

\[
\text{HR} = \frac{C_{i+1}}{C_{i} + \epsilon} \text{ and } \text{DR} = \frac{C_{i+1} - C_{i}}{C_{i+1} + C_{i} + \epsilon}
\]

\noindent 
 where $C_i$ is the number of counts in a given bin $i$ and $\epsilon = 1$ prevents division by 0. This provides insights about the overall shape of the spectrum. In addition, we observe that the distributions regularize over rounds and tend to look like gaussians. The summary statistics of the observed spectra to be fitted should also be compared with those of the simulated data to verify that their distribution does not lie at the border of the distribution of the simulated spectra. It is possible to prune the summaries for which the observation is not centred, but then the dimension of the compressed representation of the spectra may change from one round to the next, hence retraining from scratch may be required in the case of multiple round inference as detailed in App.\ref{app:pruning}.

We demonstrate the effectiveness of these summary statistics in a composite model with two overlapping components : a power law emission and an additional thermal comptonisation \citep{Titarchuk1994ApJ...434..570T}, both folded with interstellar absorption, which can be summarized as a \texttt{tbabs*(comptt+powerlaw)} in the \xspec~standard. The \texttt{comptt} model has four free parameters : the seed photon temperature (in keV), which produces a rollover at low energies, the  electron temperature (in keV) which produces a high-energy rollover, the optical depth $\tau$, and a normalisation parameter. The input values for these parameters can be found in Tab.~\ref{tab:model-parameters}. The range for the seed photon and electron temperatures are defined in such a way that they can be constrained within the somewhat limited band pass of the X-IFU instrument (0.3-12 keV, used for the range of the fit). In this section, the spectra are optimally grouped following \cite{refId0}. We fit the spectrum using 5 rounds of inference, with 5k simulations each. These parameters were chosen empirically as they represent a good compromise between speed and accuracy of inference for most spectral models with no emission lines or other high-resolution features. However, increasing both the number of simulations and the number of rounds is a simple and direct way to increase the accuracy of the posterior distribution. We compare the resulting posterior distribution with the same model fitting obtained using \texttt{BXA} with 1k live points. The corner plot in the left panel of Fig.~\ref{fig:mri-bxa} shows that the results using \texttt{SIXSA} are compatible with those obtained using \texttt{BXA}, while requiring exactly 25k convolutions with the instrument response compared to the $\sim 6$M required by nested sampling. We emphasize that unlike other approaches, \texttt{SIXSA} requires extra computing time to train the neural networks and dimensionality reduction schemes when required, but overall, we get a 10 to 100 speed up. The posterior predictive plot of the spectrum is detailed in the right panel of Fig.~\ref{fig:mri-bxa}, where we observe a good quality of the overall fit and effective constraints for the two components. We also note that the $kT$ parameter appears constrained with MRI while it is flatter using nested. This is not a question of superior performance for the MRI approach, but rather a known effect of normalising flows, which have difficulty in expressing discontinuous distributions via the transformation of Gaussian distributions \citep[see e.g.][]{rezende_variational_2016}. In Fig.~\ref{fig:summary-rounds}, we display the evolution of 4 summary statistics used in the MRI inference of Sec.~\ref{sec:summary}. The initial broad distributions shrinks as the model converges toward the best posterior distribution.

\subsection{Embedding networks}
\label{sec:embedding}

\begin{figure}
    \centering
    \includegraphics[width=\linewidth]{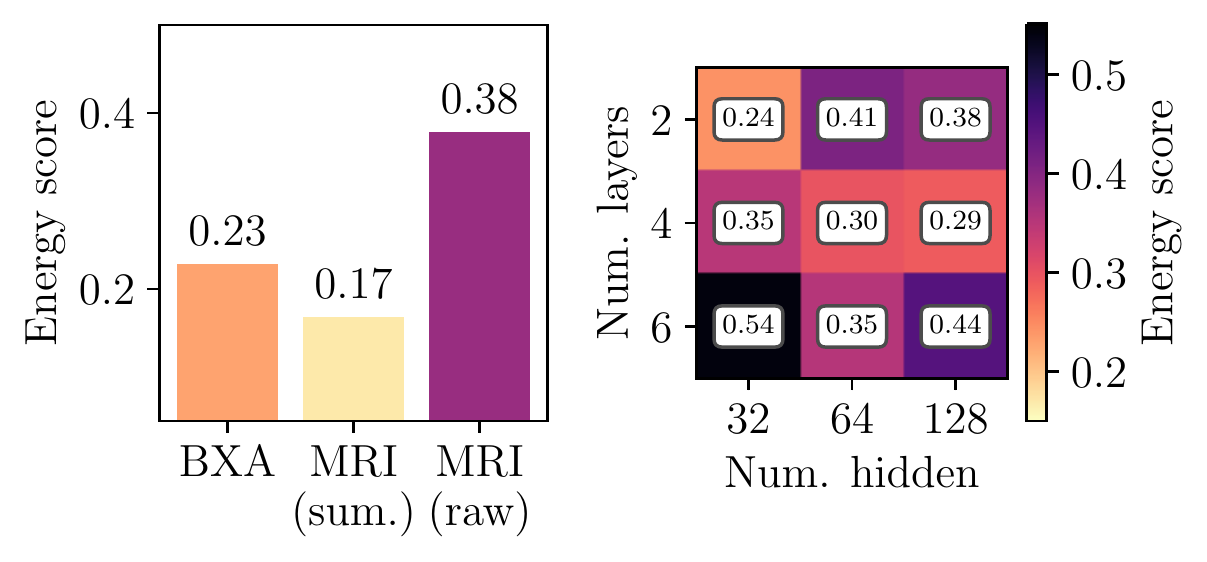}
    \caption{Energy score for posterior distributions as produced by \texttt{BXA}, MRI using summary statistics, MRI using raw spectra, and a grid of MRI using a fully connected embedding net with a varying number of layers and hidden parameters.}
    \label{fig:energy-score-embedding}
\end{figure}

An alternative approach is to learn summary statistics from simulated spectra using an embedding neural network. Instead of manually defining a dimensionality reduction scheme, we can employ an encoder architecture that automatically learns a reduction tailored to the observations we wish to constrain. Common architectures for dimensionality reduction include Multi-Layer Perceptrons (MLPs) and Convolutional Neural Networks (CNNs). In the following section, we focus on MLPs because they train quickly, whereas using CNNs for embedding significantly slows down the SBI approach—sometimes to the point where a nested sampling method becomes competitive in speed. To quantify the improvement in parameter constraints achieved by these new architectures, we use the Energy Score \citep[ES, ][]{Gneiting01032007}. The Energy Score compares the centering and spread of a multidimensional distribution against a reference parameter value. It is defined as 

\begin{figure*}[t]
    \centering
    \includegraphics[width=0.49\linewidth]{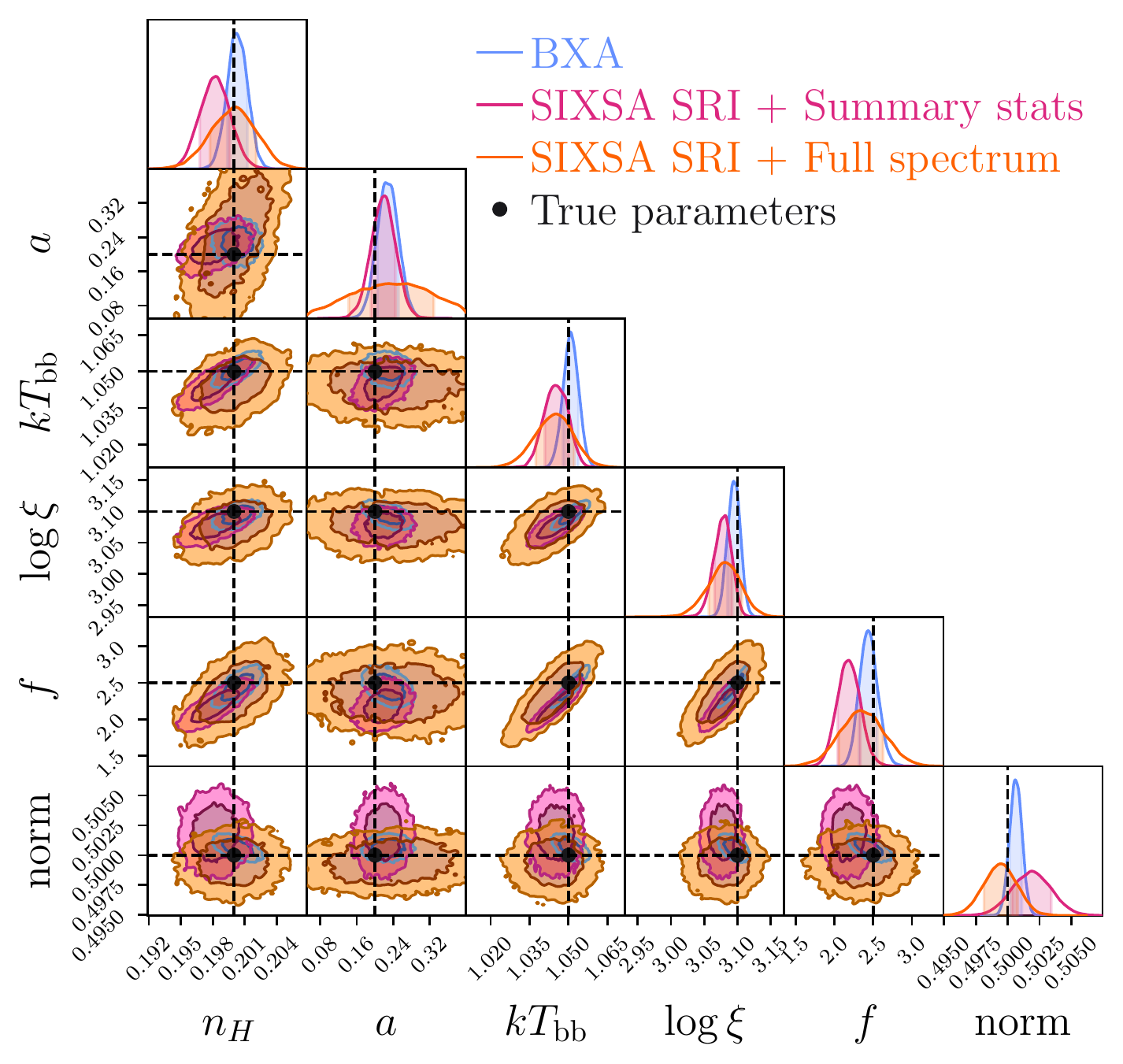}
        \includegraphics[width=0.49\linewidth]{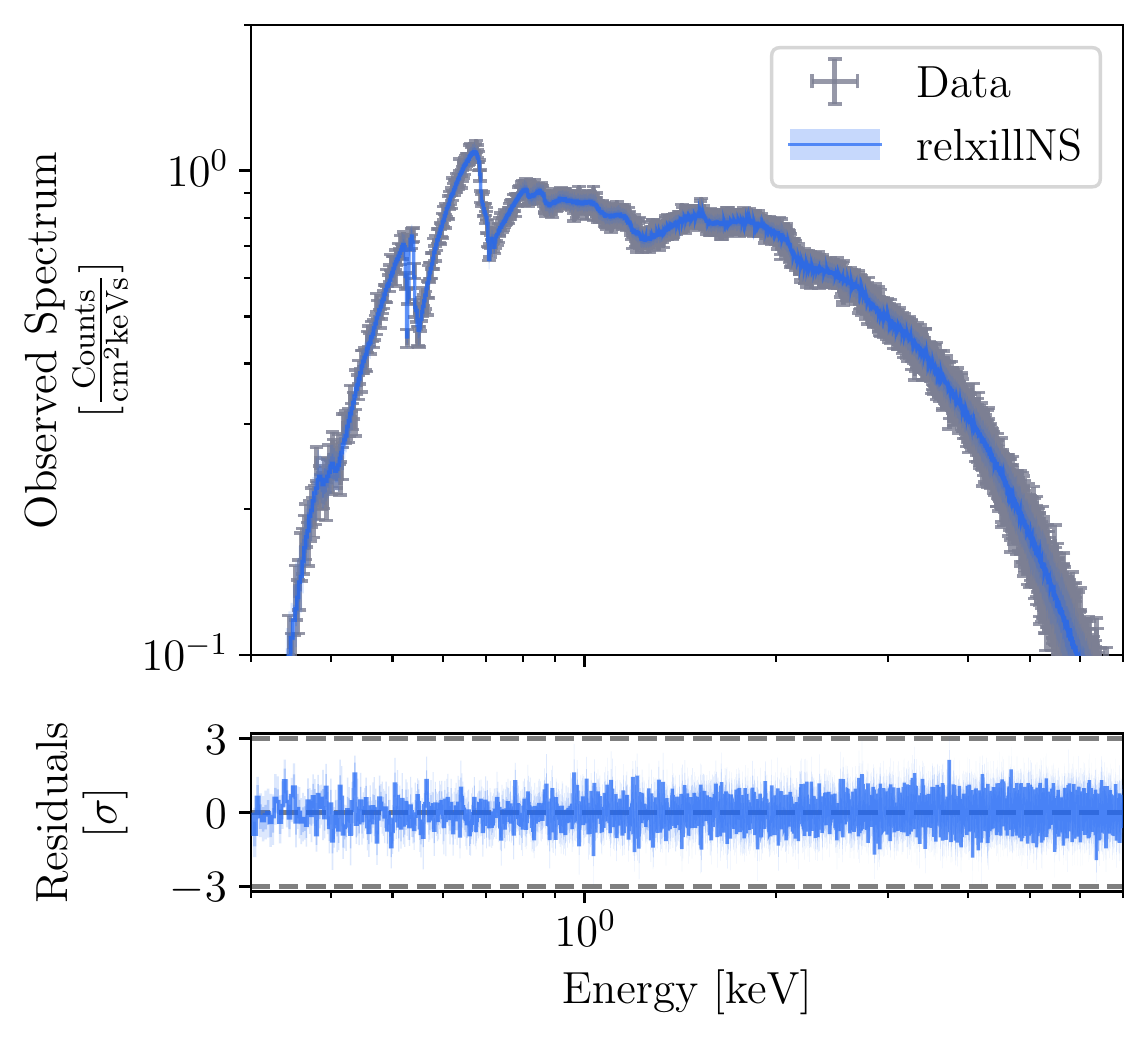}
    \caption{Left: Comparison of the posterior distributions obtained from a \texttt{BXA} run with 2k live-points and SRI run using the whole spectrum or summary stats, trained both with 200k simulations for a \texttt{tbabs*relxillNS} model. Right: Posterior predictive spectra associated with the posterior distribution obtained using summary statistics. The bands detail the (16-84) and (2.5-97.5) percentiles of the posterior predictive spectra.}
    \label{fig:sri-comparison}
\end{figure*}

$$
\text{ES}\left\{p(\boldsymbol{\theta} | \boldsymbol{x}), \hat{\boldsymbol{\theta}}\right\} = \mathbb{E} \|\boldsymbol{\theta} - \hat{\boldsymbol{\theta}} \| - \frac{1}{2} \mathbb{E} \|\boldsymbol{\theta} - \boldsymbol{\theta}' \| 
$$

\noindent where $\vec{\theta}$ and $\vec{\theta}'$ are independent random samples from the posterior distribution \( p(\boldsymbol{\theta} | \boldsymbol{x}) \) and \( \vec{\hat{\theta}} \) represents the true parameter values. A low energy score traces both a good centering of the posterior distribution with respect to the true parameter value, and also accounts for its intrinsic scatter, with lower values for more informative distributions. We use the ES as implemented by the \texttt{scoringrules} package \citep{zanetta_scoringrules_2024}. By computing the ES for various architectures while keeping \( \vec{\hat{\theta}} \) and \( \vec{x} \) constant, we can directly compare the efficiency of each approach. Using the same observational and inference setup as in Sec.~\ref{sec:summary}, MLPs with a number of layers $\in \{2, 4, 6\}$ and a number of parameters $\in \{32, 64, 128\}$ are now trained to reduce the spectra to an output of 32 dimensions in the same time as the density estimator. For each of these configurations, the ES is computed with 1k samples drawn from the surrogate posterior distributions. We display the ES for these configurations in Fig.~\ref{fig:energy-score-embedding}, in addition to the ES obtained from \texttt{BXA}. We observe that the ES obtained using embedding networks is on average worse than the ES obtained using BXA or MRI with summary statistics. This poor performance is attributed to the architecture being ill-suited to the compression task in hand. The input spectra are dense with information ($\sim 10$k channels), and scaling the MLP to fit this size would lead to a numerous number of free parameters, which would require significantly more simulations to keep up, making it more difficult to achieve relevant compression. The use of more relevant networks such as CNNs, recurrent networks and transformers is outside the scope of this paper, which is aimed at rapid and efficient inference, even if it means obtaining an approximate result. We note, however, that the exploration of more suitable methods for compressing these spectra is entirely relevant, since despite their poorer performance than simple summary statistics, inference schemes with embedding networks seem to provide results centred on the true values of our observations. Pre-trained feature extraction models suitable for X-ray spectral data could perform well in such situations, such as dedicated autoencoders or wavelet scattering transforms, which will be investigated in another paper.

On the other hand, we note that the ES is lower using MRI with summary statistics than MRI with raw spectra. This can be understood as an overfitting problem, where the neural network should learn a high dimensional space ($\sim 10$k) from a relatively small number of simulations ($\sim 20$k). Reducing the observation space to tens or hundreds of dimensions eases the learning process, and removes a lot of the redundant information that is hidden in these highly correlated observables. Finally, we note that \texttt{BXA} provides a posterior with a higher energy score compared to MRI with summary statistics. This can be related to the previous discussion in Sec.~\ref{sec:summary}, where we mentioned the smoothing of unconstrained distributions, which artificially reduces the spread of the $kT$ parameter.

\section{Feasibility study}

The science cases of \textit{Athena} \citep{Nandra_2013arXiv1306.2307N} have evolved through the reformulation of the mission to \textit{newAthena} \citep{cruiseNewAthenaMissionConcept2025}, as have the overall performances of the X-IFU, particularly in terms of instrument effective area, mostly related to a change in the mirror collection area. Tailoring observation strategies to constrain relevant physical parameters will require simulating the source in varying parameter regimes to find the best exposure possible. In this section, we demonstrate the feasibility of these studies through Simulation-based inference with X-IFU for these science cases. The unprecedented spectral resolution of X-IFU motivates the use of sophisticated physical models that allow the emission mechanisms to be described in detail. In this section, we investigate the use of SBI to perform this spectral fitting task using a relativistic reflection model, and study how well its parameters can be constrained via the sensitivity of X-IFU, within a given statistical regime (i.e. a given exposure time for a given source flux).

As discussed in Paper I, the amortization of a NN trained on a single round of simulated spectra is extremely efficient at estimating the posterior distribution for a large set of spectra. This aspect can be used to estimate the ability to measure parameters in a varying range for a given observational setup, which is extremely useful when designing an observation proposal. This is best done considering SRI, with a test sample for which we can compare the input model parameters with their reconstructed posterior distribution. Ultimately, this will tell us which model parameters X-IFU is able to constrain. Following \cite{Barret_2019A&A...628A...5B}, we are interested in checking what constraints can be derived from X-ray spectroscopy around compact objects, in particular regarding spins, but unlike them, we consider here reflection models that apply to neutron stars. For this purpose, we consider a \texttt{tbabs*relxillNS} model with \texttt{relxill v2.3.7} \citep{Garcia_2022} and the renormalize flag set to $1$ which relaxes the degeneracy between some parameters. We free five parameters of the \texttt{relxillNS} model: the dimensionless spin parameter $a$, the ionization parameter $\log \xi$, determining the ionization state of the accretion disk, the reflection fraction $f$, which describes the relative strength of reflected emission compared to the direct continuum, and the blackbody temperature $kT_{\text{bb}}$ in keV and a normalisation parameter. We generate a target observation of $\sim 3.2$M counts (roughly corresponding to a $1 \text{ mCrab}$ source observed for 100 ks) and then train a SRI inference setup with 200k simulations using the prior specified in Tab.~\ref{tab:model-parameters}. We train two inference setups, a SRI setup fed using the raw spectra, and another using summary statistics as defined in Sec.~\ref{sec:summary}. Once these setups are trained, we can infer the posterior distribution from our target observation is seconds. Note that the training process is much slower when compared to MRI setups due to the large number of simulations to iterate over. In Fig.~\ref{fig:sri-comparison}, we compare the posterior distributions from both setups to the reference posterior obtained using \texttt{BXA} with 2k live points. For this high count spectrum, we had to double the number of live points since using only 1k live points lead to overestimated uncertainties. We observe that using summary statistics yield overall better results than working directly with the full spectra, which is consistent with what we concluded in Sec.~\ref{sec:embedding}.

\begin{figure*}[t]
    \centering
    \raisebox{-0.5\height}{\includegraphics[width=0.45\linewidth]{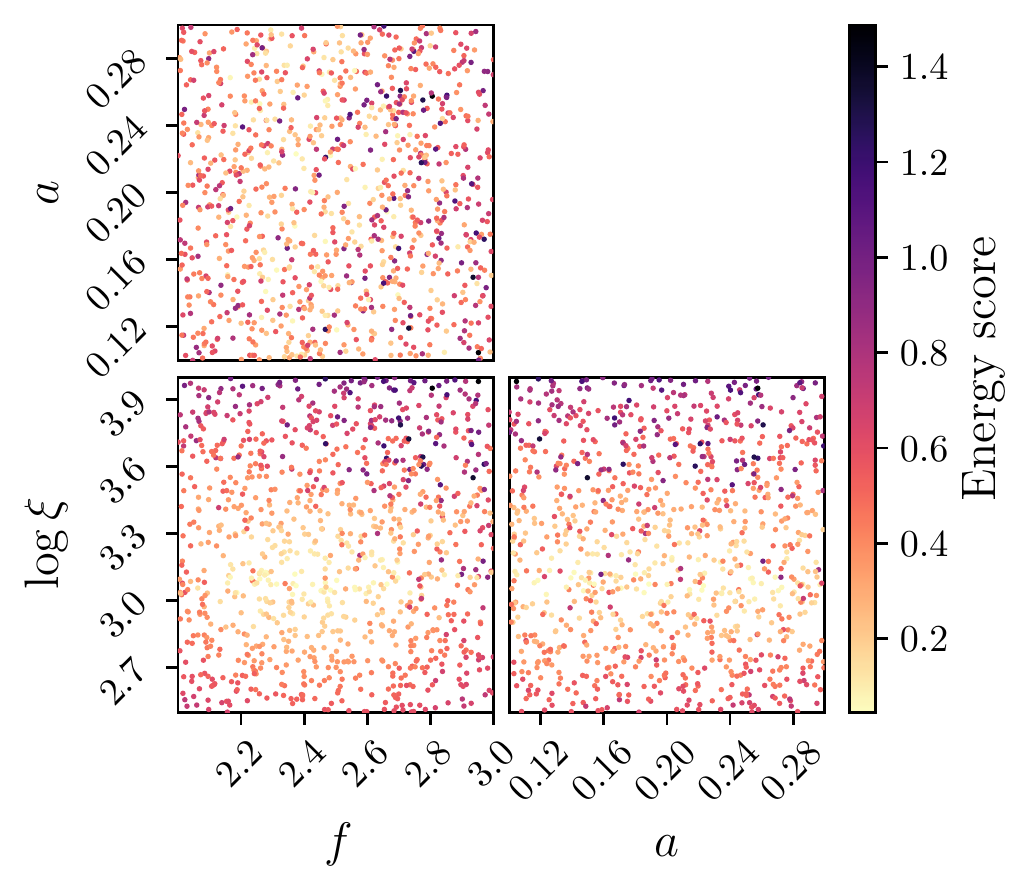}}
    \hspace{2mm}
    \raisebox{-0.5\height}{\includegraphics[width=0.5\linewidth]{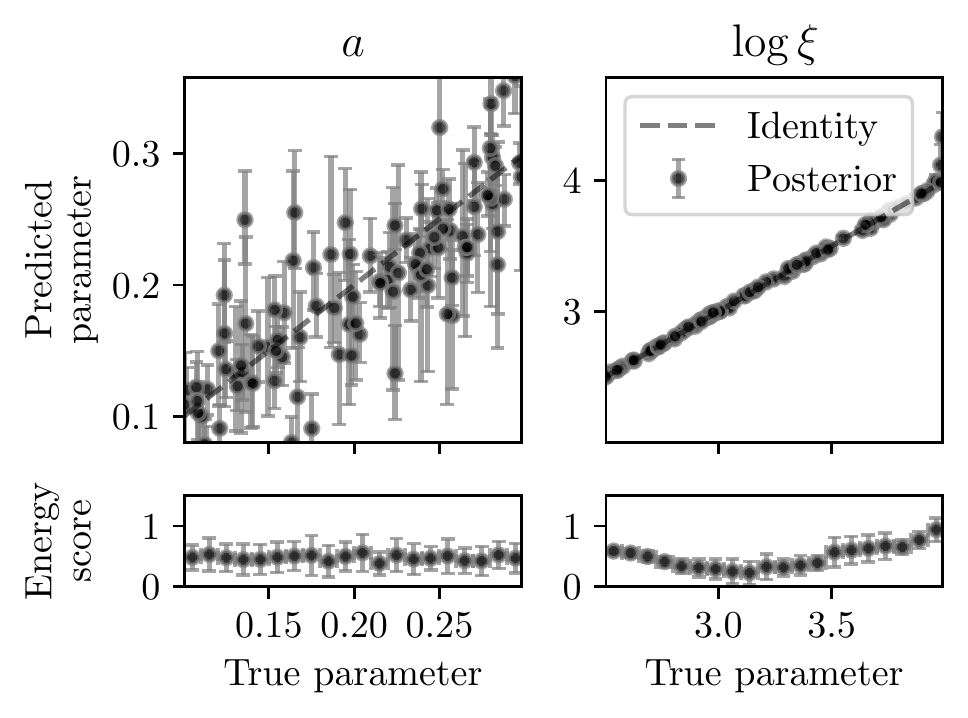}}
        
    \caption{Left : Energy score obtained from the reconstruction of the spin $a$, reflection fraction $f$ and ionisation $\log \xi$ for varying values of these parameters in a fixed observational setup. Right : Marginalized constraints for the spin $a$ and the ionisation $\log \xi$, compared to the average energy score obtained in each region of the reconstructed parameter space.}
    \label{fig:feasability}
\end{figure*}

Once the inference setup is properly trained, we can use it to generate posterior distributions for new observed spectra, generated with varying parameters around our target observation. We draw 1000 new parameter values for the spin $a$, the reflection fraction $f$ and the ionisation $\log \xi$, such that $a \sim \mathcal{U}(0.1, 0.3)$, $f \sim \mathcal{U}(2.5, 4.5)$ and $\log \xi \sim \mathcal{U}(2, 3)$. Once the posterior distribution is generated for each of these spectra, we can compute the accuracy to reconstruct the input parameters using the same Energy Score as in Sec.~\ref{sec:embedding}. In Fig.~\ref{fig:feasability}, we plot the value of ES depending on the parameters. We observe no clear trend for $a$ and $f$, where low and high ES values are distributed uniformly, meaning that the reconstruction is very noisy in the parameter range we chose. On the other hand, $\log \xi$ shows a clear trend of low ES for values $\sim 3.2$ and high ES for values $\sim 4.4$. This can be interpreted as a constraining observation for this parameter in particular, especially for $\log \xi \in [2.8, 3.5]$. We clarify this trend by marginalising the reconstructed parameters in the left panel of Fig.~\ref{fig:feasability}, where we clearly observe the ability to recover $\log \xi$ in a given range, while not being able to constrain $a$. We also note that the ES in increasing a bit for very low values of $\log \xi$, which can be attributed to the network poor performances on the boundaries of its training set.  SRI inference setup are extremely powerful at constraining a lot of spectra at once, and can unlock this kind of feasibility studies by quickly constraining portions of the parameter space. Exploring in advance our ability to constrain certain parameters of a model for a given observational setup will enable us to make the most of the new science that an instrument like X-IFU will be able to produce. We note the overall poorer constraints on the spin parameter $a$ when compared to \cite{Barret_2019A&A...628A...5B}. The consideration of a different flavour of the reflection model (\texttt{relxilllp} instead of \texttt{relxillNS}), together with a change in the instrument response, may explain why they obtained tighter constraints on the black hole spin parameter in their study, compare to the poor constraints we get on the neutron star spin parameter. It remains that within this statistical regime, probing shorter timescales, tighter constraints can be derived for other parameters, e.g. the ionization parameter. It is worth adding that multi-layer coatings is currently envisaged for the \textit{newAthena} optics, to boost the effective area of the mirror around the Iron K line, and this should help to get better constraints on the neutron star spin parameter from reflection spectroscopy.

As an additional discussion, we note that the performance of simulation-based inference in cases of high signal-to-noise ratio is reduced. The normalising flows take advantage of the observation noise to create a mapping between the parameters and the summary statistics that works on average. As with exact approaches, as the signal-to-noise ratio increases, the volume of a posteriori parameter distributions decreases, making it more difficult to produce an exhaustive mapping of this region. Nevertheless, it is possible to couple this feasibility approach with a priori distribution restrictions, as in Paper I. In this situation, for example, the network could be trained more finely to reconstruct spectra between 2 and 4 million counts, thus facilitating constraints on the parameters. 

\section{Two temperature plasmas with X-IFU}
\label{sec:apec}

The ICM can present multiple phases due to dynamical interactions linked, e.g. to the presence of a cool-core, or strong feedback from the AGN in the brightest galaxy of the cluster. Probing such spectra can be tricky because of the high degeneracy between the two similar components. In particular, suing SBI with the previous summary statistics, we expect difficulties in assessing the redshift and velocities, which are features encoded in the spectral line position and width. To explore this, we consider a \texttt{tbabs*(bapec+bapec)} model in \xspec\ terminology with parameters highlighted in Tab.~\ref{tab:model-parameters}. The Astrophysical Plasma Emission Code \citep[APEC,][]{smith_collisional_2001} simulates the X-ray emission of an ionized gas, including the continuum mainly dominated by the \textit{bremsstrahlung}, and emission lines. The parameter $kT$ controls the gas temperature in keV and the abundance $Z$ is linked to the metal content of the gas. A mock observation with a target number of counts of $\lesssim 300 k$ is generated (which corresponds to counts generated by the full field of view of a $\sim 750 \text{s}$ integration in the Perseus core). 

\begin{table}[]
    \caption{Line complexes injected as extra summary statistics for the inference of the \texttt{tbabs*(bapec+bapec)} model.}
    \label{tab:complexes}
    \centering
    \begin{tabular}{llll}
    \hline
    Complex & $E_\text{low}$ [keV] & $E_\text{high}$ [keV] & $N_\text{bins}$ \\
    \hline
    \hline
    O VIII & 0.62 & 0.66 & 20 \\
    Fe XXIV + Ne X & 0.97 & 1.2 & 20 \\ 
    Fe XXIV + Mg XII & 1.4 & 1.55 & 20 \\ 
    Fe XXV & 6.4 & 6.65 & 20 \\ 
    \end{tabular}

\end{table}

\begin{figure*}
\centering
    \subfigure[O VIII]{
    \includegraphics[width=0.45\hsize]{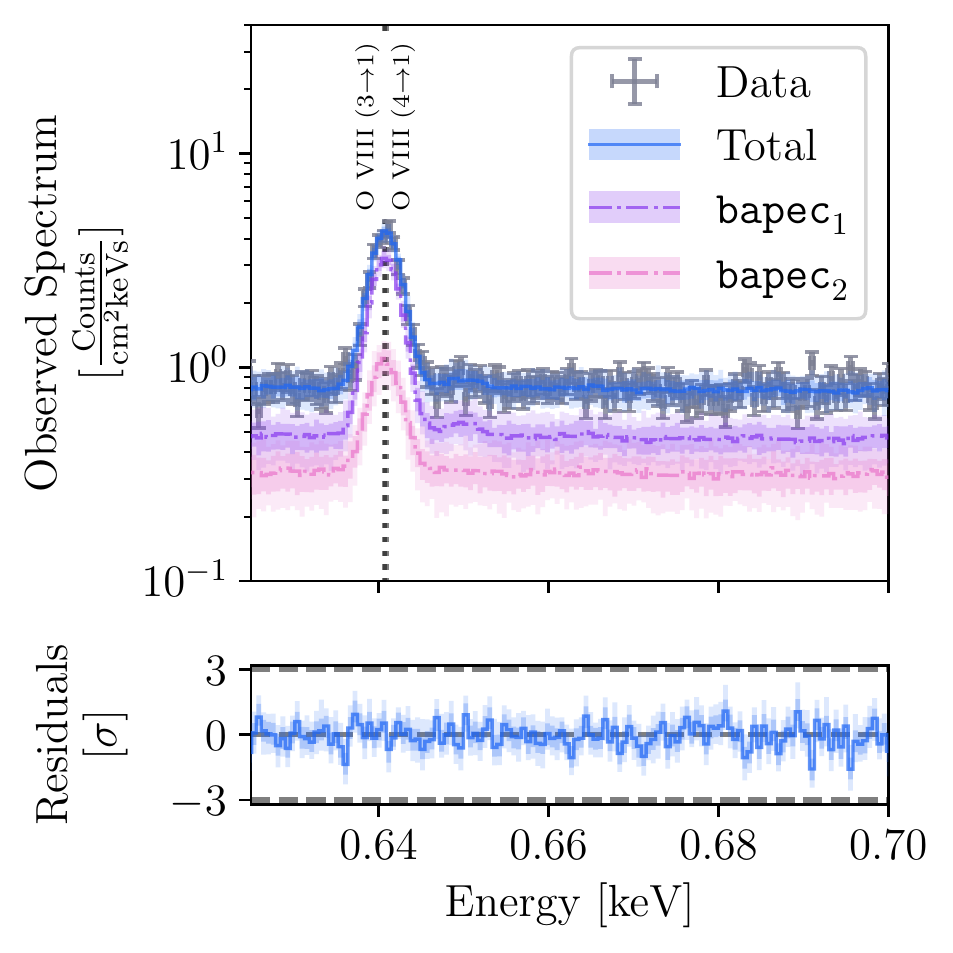}
    \label{fig:o-complex}
    }
    \subfigure[Fe XXIV + Ne X]{
    \includegraphics[width=0.45\hsize]{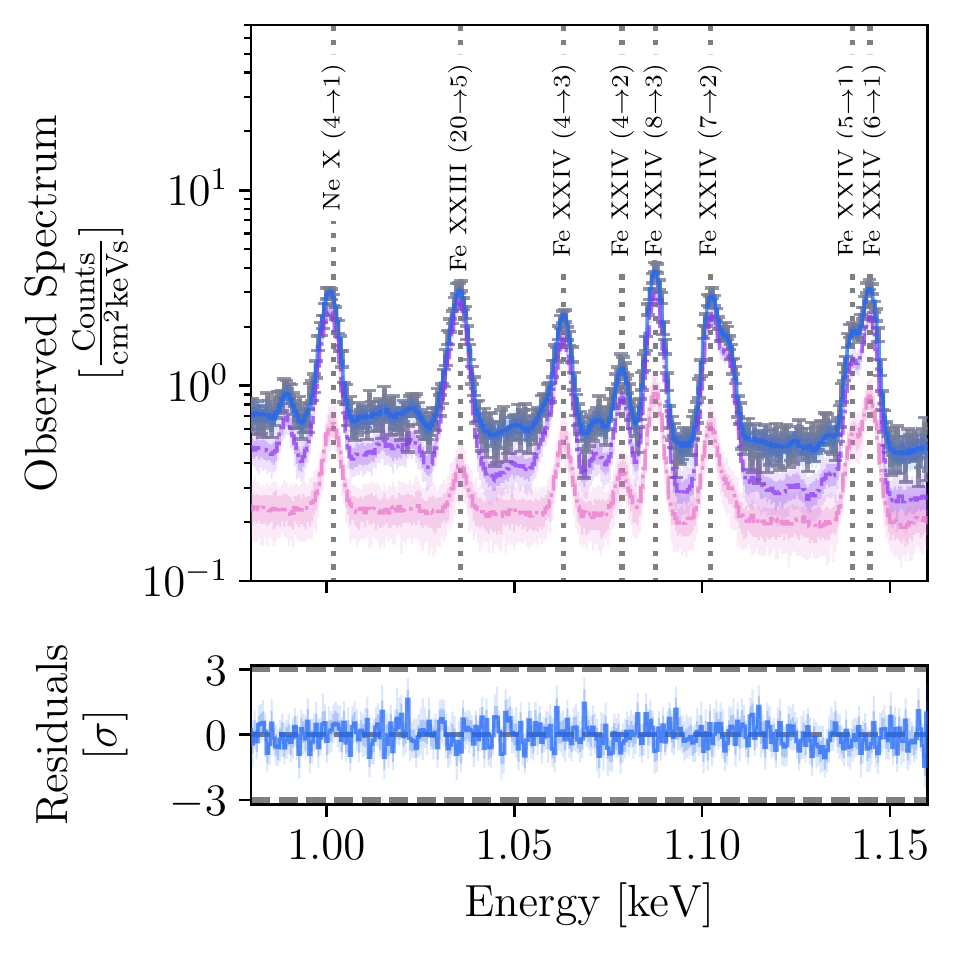}
    \label{fig:fe-ne-complex}
    }
    \subfigure[Fe XXIV + Mg XII]{
    \includegraphics[width=0.45\hsize]{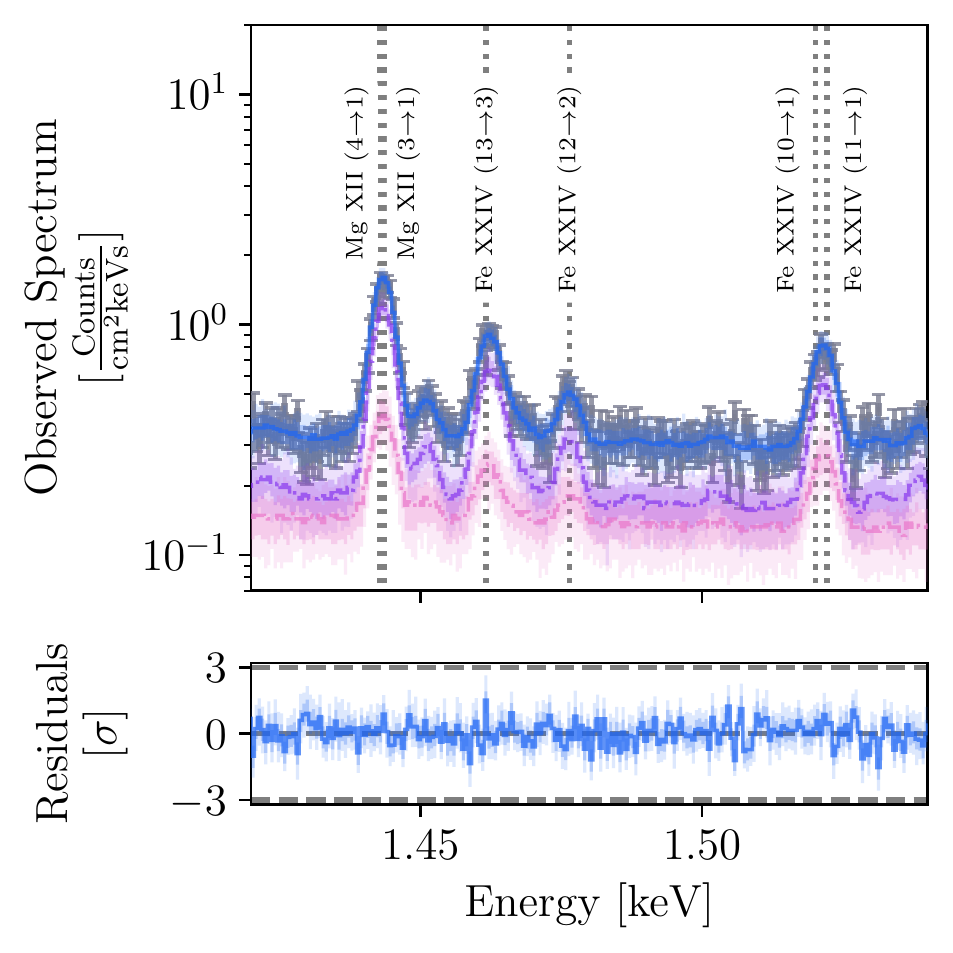}
    \label{fig:fe-mg-complex}
    }
    \subfigure[Fe XXV]{
    \includegraphics[width=0.45\hsize]{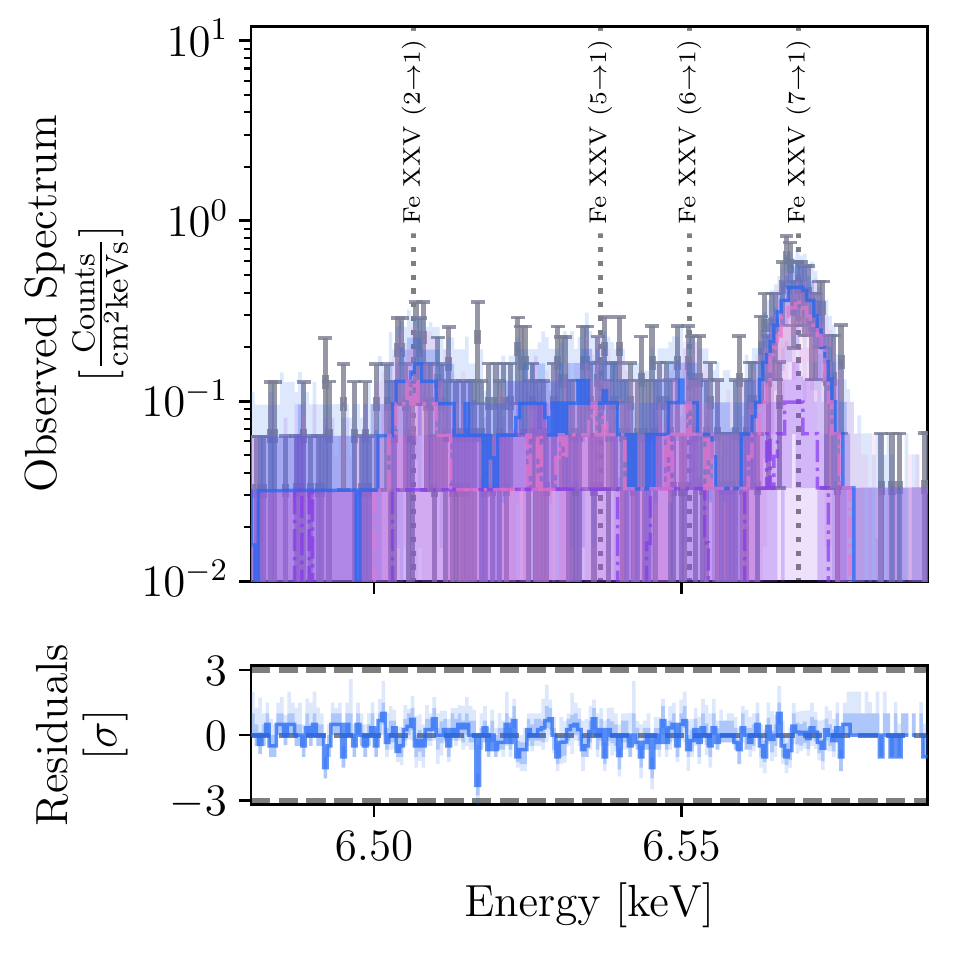}
    \label{fig:fe-complex}
    }
    \caption{Posterior predictive spectra of the 4 line complexes added as extra summary statistics for a MRI run with a \texttt{tbabs*(bapec+bapec)} model for 20 rounds of 10k simulations. The brightest emission lines are highlighted in each plot.}
    \label{fig:apec-complexes}
\end{figure*}

Summary statistics as defined in Sec.~\ref{sec:summary} remove the information associated with the spectral lines by diluting them in large energy bins. To overcome this, we add a specific focus on some line complexes, associated with energy bands detailed in Tab.~\ref{tab:complexes} for which we subdivide the counts in 20 smaller bins. We compute the average energy $\bar{E}_i$ in each bin $i$, weighted by the number of counts: 

\[
\bar{E}_i = \sum_j E_j C_j / \sum_j C_j
\]

\noindent where the summation of $j$ represents the summation of the instrumental channels within the bin $i$, and $E_j$ is the mean energy of the channel. This extra statistic can grasp line-related information, such as their shift and broadening, which are directly related to the redshift $z$ and velocity $v$ parametrized by the \texttt{bapec} model. This is further discussed in Sec.~\ref{sec:interpretability}. Due to a much larger number of summaries, we expect a slower convergence of the neural network. Hence, we fit this spectrum using MRI with 20 rounds of 10k simulations. In Fig.~\ref{fig:apec-comparison}, we display the full posterior distribution compared to a \texttt{BXA} run with 1k live points, along with the posterior predictive spectrum in the [0.5-8] keV band. We observe a good agreement between both posterior distributions, with a more significant spread in the \texttt{SIXSA} reconstructed parameters. As expected, when working with compressed representation for a model with high-frequency features such as emission lines, we lose information on the redshift and velocity, leading to broader distributions for these two parameters. As shown in Fig.~\ref{fig:apec-complexes}, we get a satisfactory result for the 4 line complexes we focused on during the inference, where emission lines are well constrained even if the posterior parameter distribution are not optimal.  

\begin{figure*}
    \centering
    \includegraphics[width=0.45\linewidth]{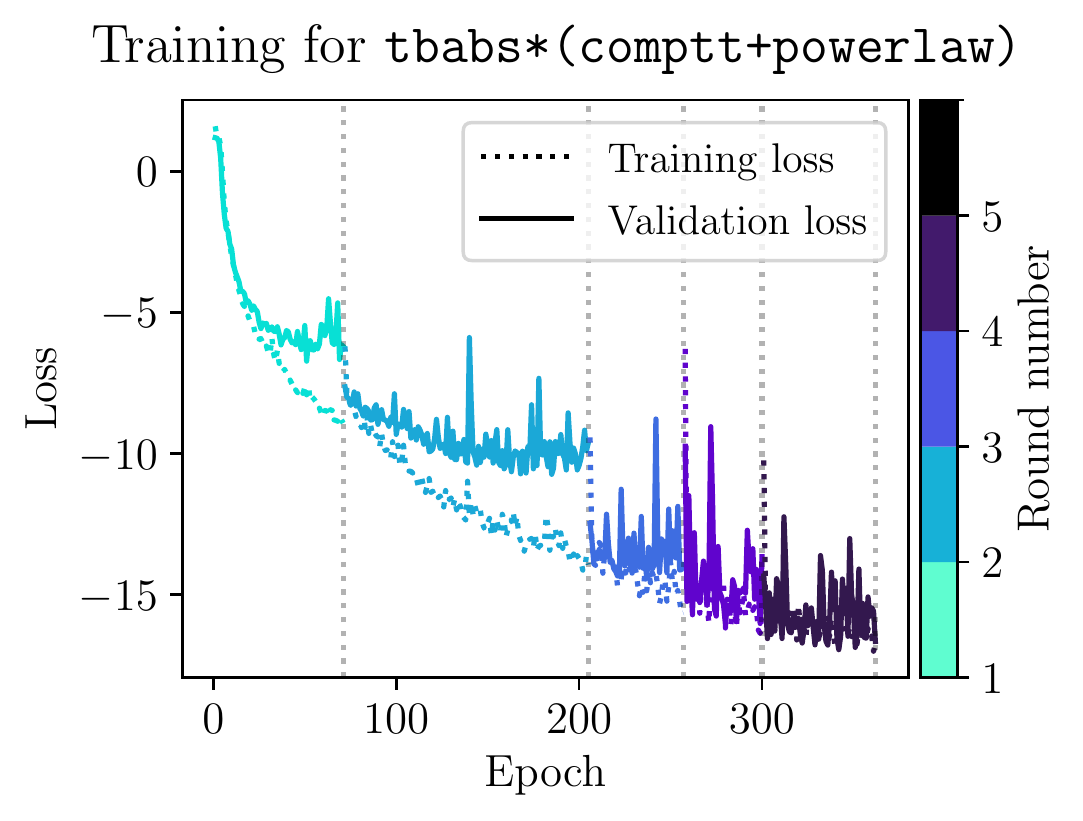}
    \includegraphics[width=0.45\linewidth]{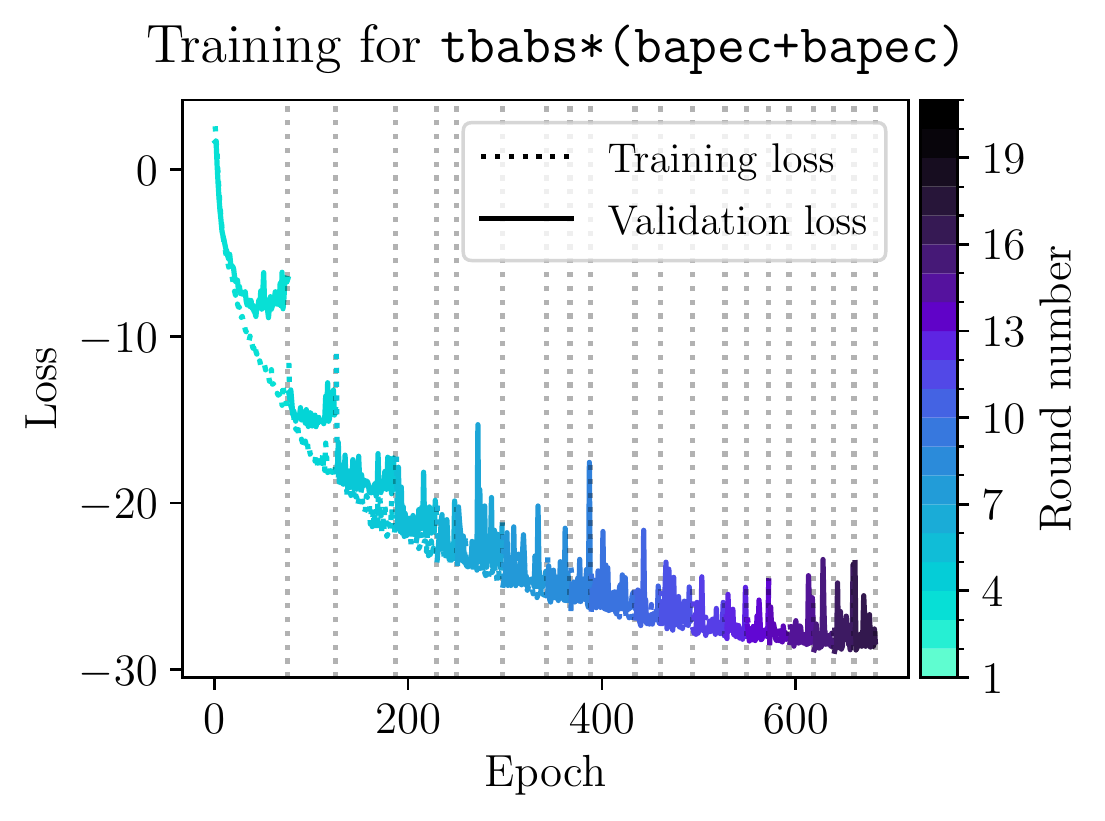}
    \caption{Evolution of the loss for the training and validation set depending on the training epochs and inference round for two spectral models used in this paper. Each round adds a set of 5k simulations to the global set for the left panel and a set of 10k simulations for the right panel.}
    \label{fig:training-curves}
\end{figure*}

We emphasize the necessity for adding emission lines tailored summary statistics in Fig.~\ref{fig:apec-full-comparison}, where we observe a significant offset and uncertainty on the redshift and velocity parameters when using only sumarry statistics without focus on the line complex structures. We believe that it is possible to further improve the characterisation of these spectra by investigating summary statistics that better represent this information from the emission lines. Fourier space approaches, wavelet transforms or other methods based on multiscale decomposition such as wavelet scattering transforms \citep[e.g.][]{JMLR:v21:19-047} could improve the quality of the results. However, such investigations are beyond the scope of this paper.

\section{Discussion}

\subsection{Assessing convergence}

At first sight, convergence can be hard to assess for SBI approaches. How many rounds and simulations are needed to get a proper and stable result? A way to tackle with this is to inspect the training curves of the neural network. In Fig.~\ref{fig:training-curves}, we display the evolution of the loss evaluated on the training set and validation set. In the first rounds, we observe a rising validation loss compared to a decreasing training loss, which is a signature of early overfitting. In the first rounds, the training set is too small to let the network learn the full dynamic of the mapping between the parameters $\boldsymbol{\theta}$ and the observables $\boldsymbol{x}$. As the number of rounds increases, the total number of simulations and the simulations are closer to the best-fit value, which stabilizes the training. Increasing the number of simulations can prevent overfitting, as single round inference is feasible. However, it requires much more simulations. In general, there is no given rule for stating that a training is over, and the number of epochs and rounds can be considered as another hyperparameter of the network. However, some metrics can give ideas about the quality of the training. To check whether the inference setup converged or not, we can check that the training and validation loss are not improving significantly when running extra inference round. We can also check for the stability of the surrogate posterior distributions, which should concentrate toward its most informative state when the number of rounds increases. 

\subsection{Are we aiming at the best-fit?}

\begin{figure}[t]
    \centering
    \includegraphics[width=\linewidth]{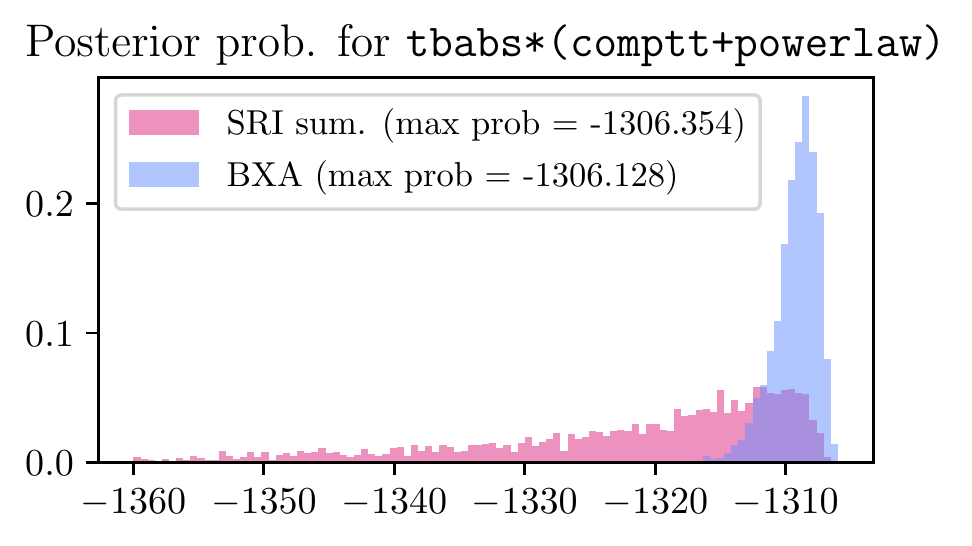}
    \includegraphics[width=\linewidth]{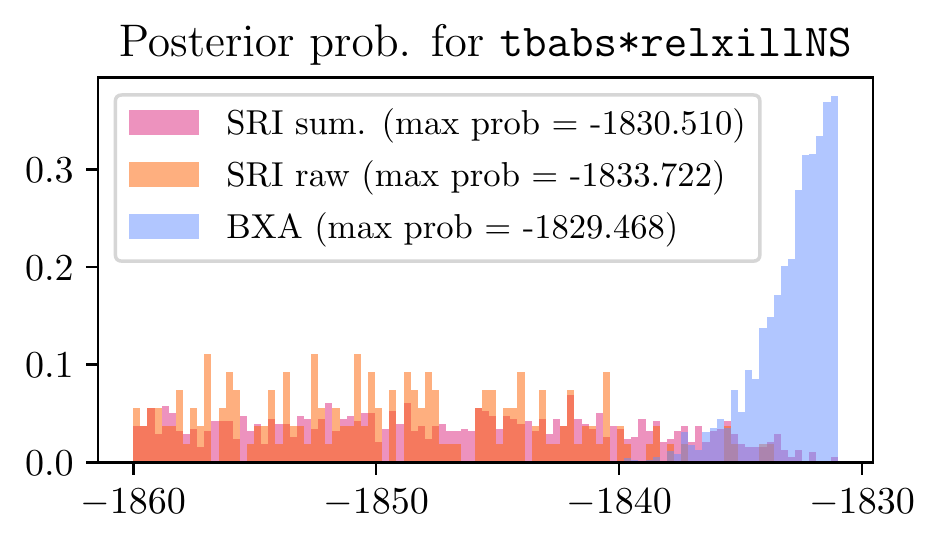}
    \includegraphics[width=\linewidth]{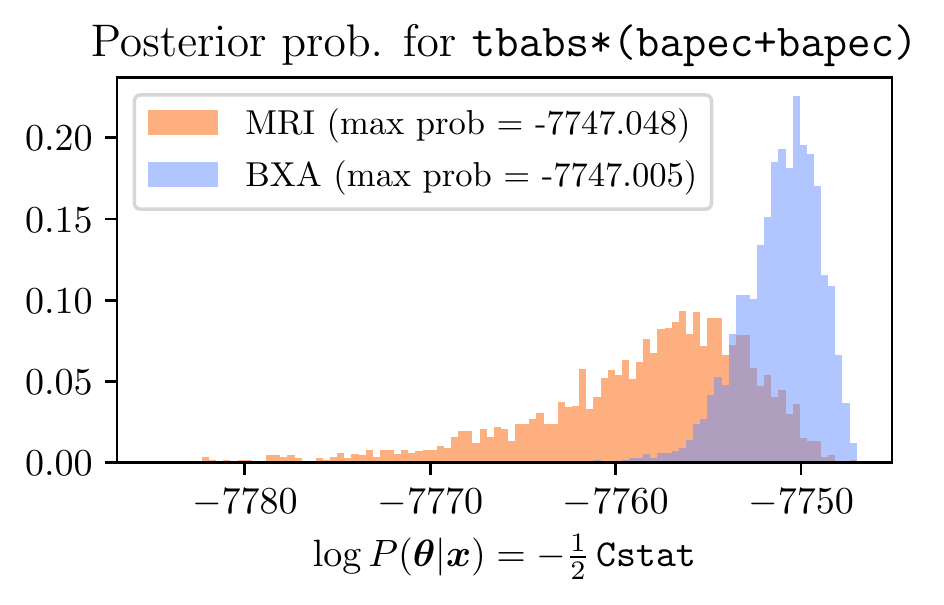}
    \caption{Comparison of the distribution of posterior probability for the samples produced using BXA and the various SBI inference pipelines used in this paper. }
    \label{fig:cstat-comp}
\end{figure}

From the surrogate posterior distributions produced with SBI inference pipelines, we can generate a distribution of posterior log-probability associated with each point, as evaluated in the same standard as \texttt{BXA}. This quantity is directly linked to the well known \texttt{Cstat} such that $\log P(\boldsymbol{\theta}|\boldsymbol{x}) = -\texttt{Cstat} /2$. The posterior probability distributions for the 3 problems we treated in this paper are displayed in Fig.~\ref{fig:cstat-comp}. For each of these inference problems, we observe that the probability distribution is wider and lower on average compared to \texttt{BXA}, meaning that the surrogate posterior distributions include parameter values that are worse than the one obtained using exact techniques. This translates either the small offsets or the larger spread that can be observed in the posterior comparisons, such as in  Fig.~\ref{fig:mri-bxa}, Fig.~\ref{fig:sri-comparison} or Fig.~\ref{fig:apec-comparison}. Using SBI approaches with multi-round inference, we can obtain a fairly good approximation of the best fit, within a 10 percent of the best fit as computed with exact methods. With these approximated distributions, it is hard to directly obtain the best-fit per se, but, we can use them to initialise exact methods to increase their efficiency, we can either start a traditional minimisation from the best-fit value obtained using the SBI approach, or use the surrogate distributions as proposal distribution for increasing the convergence speed of MCMC methods. Such synergies could improve both the speed and reliability of exact methods, thanks to the fast convergence and robustness of SBI. 

\subsection{Understanding the summary statistics}
\label{sec:interpretability}

As stated before, the choice of summary statistics for a given problem is not unique, and different choices can lead to different posterior distributions. In general, the choice of summary statistics falls under the hyperparameter optimisation, and is not likely to be optimised if a working inference setting is found for a given problem. Instead of defining this set a priori of our problem, we motivate a posteriori the choice of summaries we performed in this paper, and in particular, how the addition of the weighted energy in the two components plasma model of Sec.~\ref{sec:apec} allowed to constrain properly the redshift with spectral lines information. To investigate the sensitivity of our summary statistics to a given parameter, we rely on identifying active subspaces, as proposed by \cite{constantineActiveSubspaceMethods2014}. Active subspace aims at identifying covariant directions in a multivariate function to define a preferred and smaller parameter space. However, we are not interested in finding such a space, but rather by identifying the parameters that are sensitive to summary statistics, so we get an interpretable view of what kind of information and constraints they bring. To do so, we first define an arbitrary transformation between the parameter space $\boldsymbol{\theta}$ and a given statistic $x_i$. By fitting a lightweight neural network \citep[a two layer \texttt{ResNet} with 100 hidden features, see][]{he2015deep} on simulated samples, we obtain a transformation $f$ which predicts the expected value of $x_i$ given a set of parameters $\boldsymbol{\theta}$. As it is parametrized using neural network, it is easy to derivate, and in particular, we can compute the sensitivity matrix $M$ defined as 

$$
M = \mathbb{E}_{p(\mathbf{\theta}|x_i)}\left[ (\nabla_{\boldsymbol{\theta}}f). (\nabla_{\boldsymbol{\theta}}f) ^T\right]
$$

\noindent which is the expected value of the outer product of the gradient of $f$ estimated on the posterior distribution support. $M$ conceptually relates to the uncentred covariance of the gradient function, and the Fisher information if $f$ is a log-likelihood. The eigenvalues and eigenvector of this sensitivity matrix trace directions of high correlation between the parameters and the investigated summary statistic. Computing the sum of the eigenvectors weighted by the eigenvalues in the parameter space gives the overall preferred direction of the parameters for this given summary statistics, which we further refer to as the sensitivity. 

\begin{figure}[t]
    \centering
    \includegraphics[width=\linewidth]{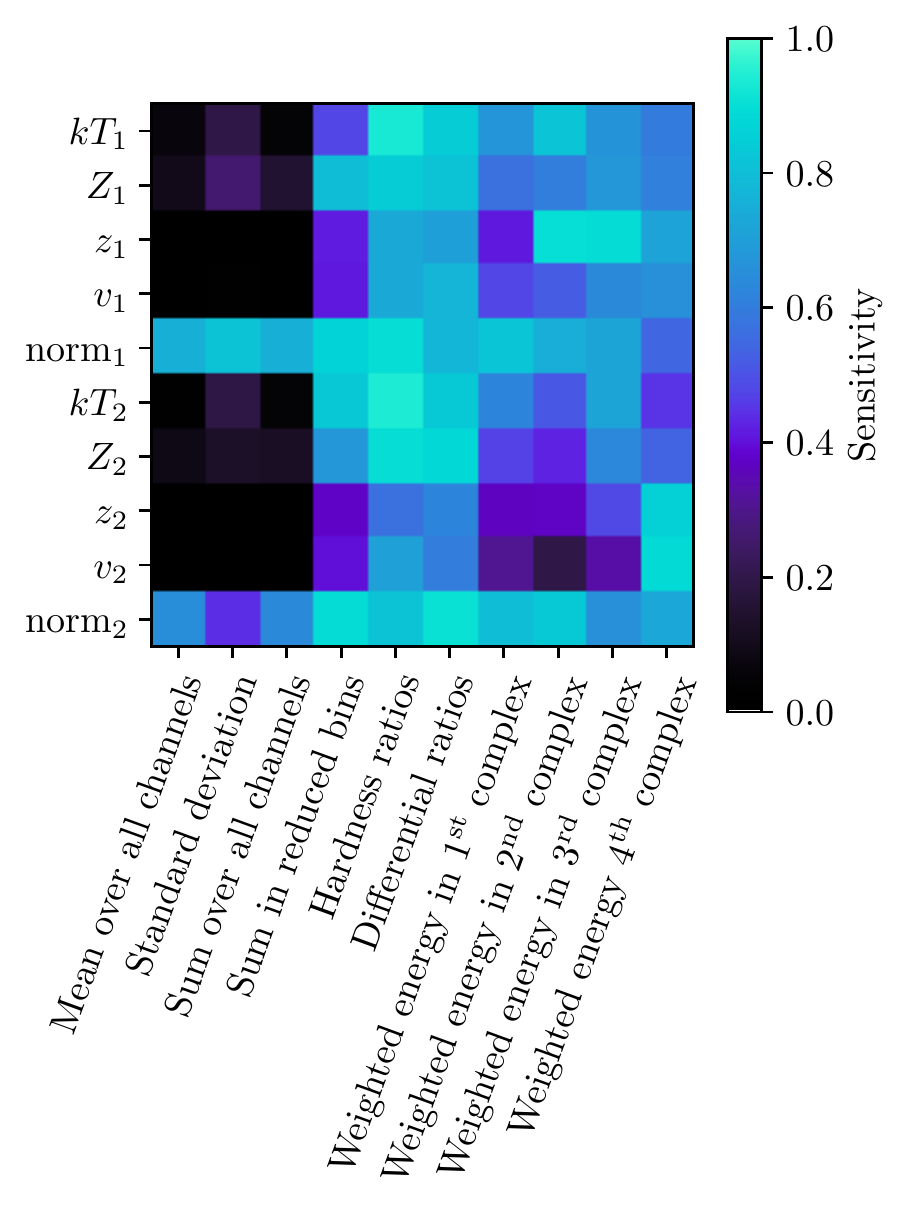}
    \caption{Sensitivity of each summary statistics to the parameters of a \texttt{tbabs*(bapec+bapec)}. The sensitivity arrays are normalized to allow qualitative comparison between the summary statistics.}
    \label{fig:apec-sensitivity}
\end{figure}

We apply this method to the double plasma model from Sec.~\ref{sec:apec}, and investigate the link between the statistics and the model parameters. We use standard normalized parameters to facilitate the training of $f$ for each statistic, and reduce the sums, hardness ratios, differential ratios and weighted energies by taking the maximum sensitivity for each band considered. The resulting sensitivity is further normalized to get comparable quantities for each summary statistics. The corresponding sensitivities are displayed in Fig.~\ref{fig:apec-sensitivity}. From this figure, we can understand how each summary helps us in constraining a specific parameter. For instance, global quantities such as the mean, sum and standard deviation are mostly sensitive to the normalisation of both components, while statistics computed in bins such as the sum, hardness ratios and differential ratios trace the shape of the spectrum with their high correlation to the temperature. Finally, the weighted energy computed in complexes informs about the redshift and velocity. As shown in Fig.~\ref{fig:apec-complexes}, the first component dominates the first three complexes, showing good correlation of the redshift $z_1$ for the second and third. On the other hand, the last complex is dominated by the second component, which shows good correlation for both $z_2$ and $v_2$. The increasing sensitivity to the velocity with the complex number can point to its increased effect with energy, as the induced broadening scales as $\sigma \propto vE$. Such an analysis further highlights the need for additional statistics to constrain the line-related information in this model and brings interpretability to the SBI approach with physically motivated compression, unlike principal component analysis. It also illustrates how incorporating specific domain knowledge is key in improving the performance of ML approaches. 

\section{Conclusions}

In tpaper,aper we demonstrate the feasibility of simulation-based inference for high-resolution X-ray spectroscopy. Simulation-based inference relies on neural networks learning probability density transformations between the parameter space and the simulated observables. Without the need to explicitly formulate a likelihood for the problem, it can be used to obtain well-defined posterior distributions for the parameters. In contrast to our previous work, we focus here on high-energy resolution spectra such as those produced by \textit{newAthena}/X-IFU. The key to using SBI approaches for high-resolution spectroscopy is spectral compression using summary statistics, which reduces 24k channel spectra to a hundred or so values that encapsulate their spectral features. We show that this approach allows us to recover the parameters of multi-component models as well as exact computation using nested sampling, while drastically reducing inference time and the number of likelihood evaluations required. We also explore the use of embedding networks to automatically learn compression schemes, but they show poor performance compared to hand-crafted summary statistics. Single-round inference requires many simulations to train an amortized network, but once trained it can infer parameter values almost instantaneously. It can be used to fit many spectra and, for example, explore the constrainable parameter values for a given observation proposal. Multiple-round inference can only fit a single spectrum, but it provides well-calibrated posterior distributions with significantly less computational effort and time compared to exact methods. The use of summary statistics on top of these two approaches increases the interpretability of this machine learning approach, as specific features such as emission lines require specific summary statistics that are physically motivated. We believe that SBI has great potential for the next decades of high-energy astrophysics, as it drastically reduces the computational requirements while still being close to exact inference. Finding efficient compression schemes is the key to improving this approach and fully exploiting the amazing data that the coming high-resolution X-ray spectrometers will bring us.

\begin{acknowledgements}
SD \& DB would like to thank all their colleagues, in particular Fabio Acero, Alexei Molin, Etienne Pointecouteau for their support and encouragement in developing further the power of SBI-NPE. Special thanks to Jonas Gérard and Yannis Guillelmet for contributing to the testing of the \texttt{sbi} toolkit. We also thank Thomas Dauser for support in implementing and testing extensively the \texttt{relxill} package, whose models are part of the most challenging ones tested here. In addition to the \texttt{sbi} package \citep{Tejero2020JOSS....5.2505T}, this work made use of many awesome Python packages: \texttt{ChainConsumer} \citep{Hinton2016}, \texttt{matplotlib} \citep{matplotlib}, \texttt{numpy} \citep{numpy}, \texttt{pandas} \citep{pandas}, \texttt{pytorch} \citep{pytorch}, \texttt{scikit-learn} \citep{scikit-learn}, \texttt{scypi} \citep{scipy}, \texttt{tensorflow} \citep{tensorflow}.
\end{acknowledgements}

\bibliographystyle{aa} 
\bibliography{dbarret} 

\appendix
\onecolumn
\section{Including the background}
\label{app:bkg}
Including a background spectrum in the inference using SBI is straightforward. If an independent spectrum containing a background realisation is available, and if a motivated model exists for it, it is only necessary to add another spectral component to the spectral model, which is fitted to the background spectrum and the source+background spectrum, in addition to the source spectral model. However, in this section, we chose to focus on the capability of SBI to perform automatic marginalisation over nuisance parameters. If we have an independent realisation of the background, we can add a Poisson realisation of it to the modelled source spectrum during the simulation process. Accounting for such an effect in a traditional likelihood framework would require to perform marginalisation over the background parameters, thus involving numerical integration over each bin of the spectrum. Using SBI, this is straightforward to implement and requires almost no computational overhead. 

\begin{figure*}[h!]
    \centering
    \includegraphics[width=0.49\linewidth]{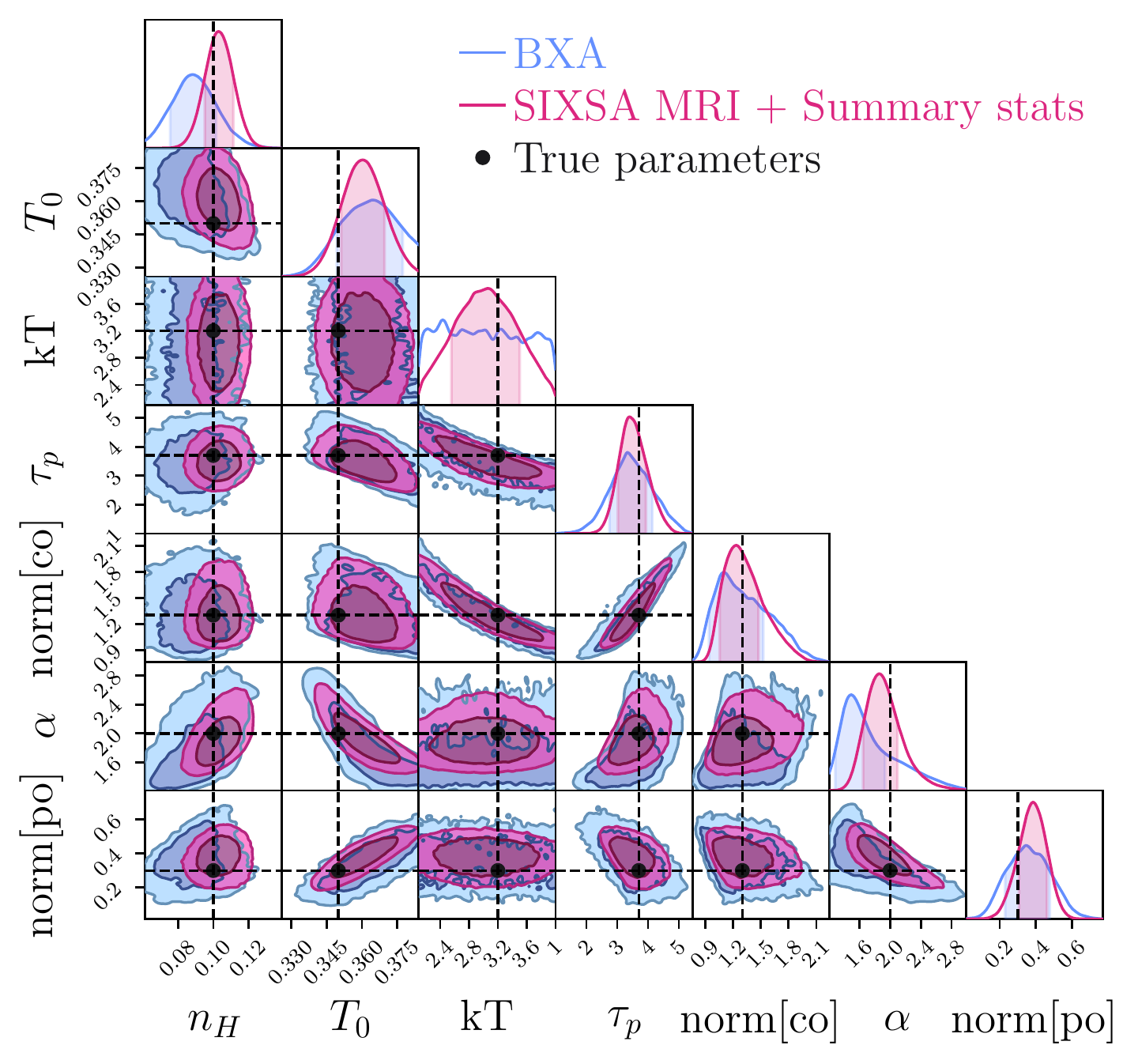}
    \includegraphics[width=0.49\linewidth]{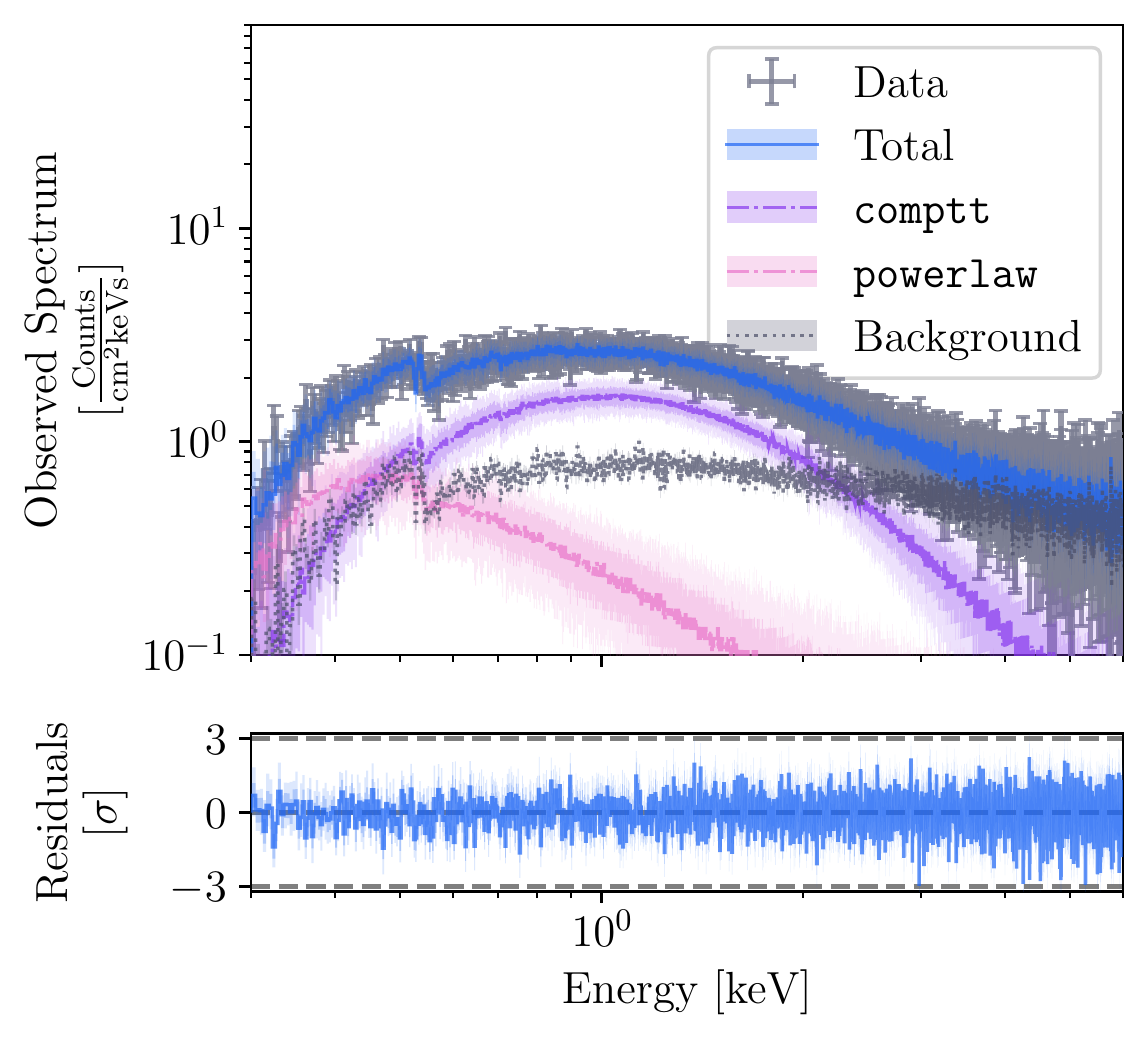}
    \caption{Left: Comparison of the posterior distributions obtained from a \texttt{BXA} run with 1k live-points and a MRI run with 5 rounds and 5k simulations each using summary statistics for a \texttt{tbabs*(comptt + powerlaw)} model, with the inclusion of a background spectrum. Right: Posterior predictive spectra associated with the posterior distribution obtained using summary statistics. The bands detail the (16-84) and (2.5-97.5) percentiles of the posterior predictive spectra.}
    \label{fig:bkg-bxa}
\end{figure*}

We demonstrate this using a similar observational setup as in Sec.~\ref{sec:summary}, considering the case where a background spectrum is provided independently. We generate two background spectra, one to be added to the source spectrum and another independent realisation that we consider as our observed background. Inference is performed on the source+background spectrum. The background is simulated using a \texttt{tbabs*powerlaw} model parameterised to contribute $\sim 10\%$ of the total counts of the source+background spectrum. The results are compared to a \texttt{BXA} run with 1k livepoints with \texttt{cstat} minimisation, and the posterior parameters are displayed in Fig.~\ref{fig:bkg-bxa}, where we observe a good agreement between the two approaches, while remaining orders of magnitude faster to compute. 

\newpage

\section{To prune or not to prune}
\label{app:pruning}

We have shown in this paper that the regularisation of the observables using compression schemes such as summary statistics is a key feature in improving the efficiency of the SBI approach. In particular, the centrality of the observable in the training set induced by the summary statistics increases the training efficiency. In this sense, it is easy to prune summary statistics that do not cover the spectrum statistics. This is expected to increase both the speed and quality of training, as we work with fewer and more representative statistics. However, when pruning in a multi-round scheme, as the simulations get closer to the observable after each round, some statistics become relevant again, while others should be discarded. In this situation, we have to retrain the network from scratch in each round, as the nature of the observable changes. In addition, we cannot train on the previously generated simulations, as they are no longer relevant. So when pruning, we have to consider this trade-off between speed and efficiency. Fig.~{fig:pruning-training} shows how the training and validation losses evolve when the network is retrained from scratch. This demonstrates the network ability to converge in a small number of epochs, thanks to the proposal distribution generating a training sample that is closer to the true posterior distribution with each iteration.

\begin{figure}[h!]
    \centering
    \includegraphics[width=0.5\linewidth]{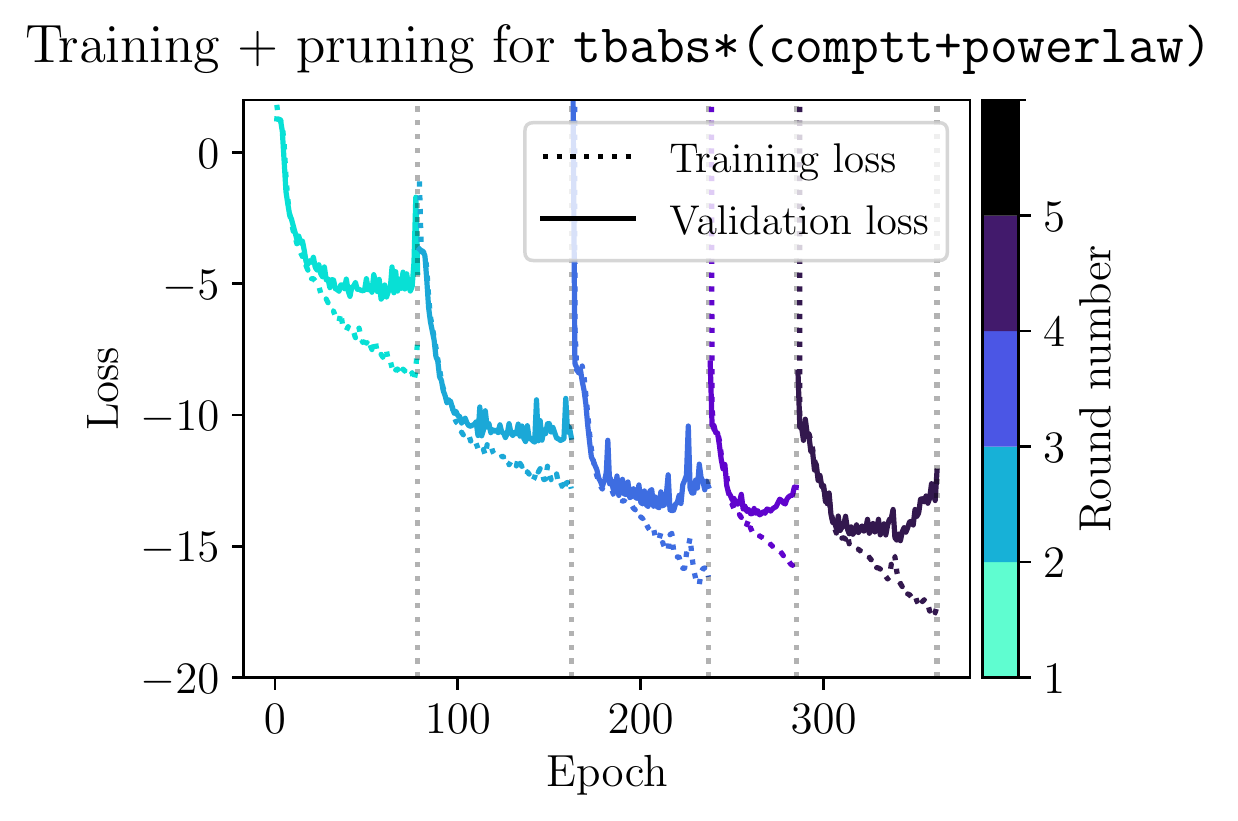}
    \caption{Evolution of the loss for the training and validation set depending on the training epochs and inference round for the \texttt{tbabs*(comptt+powerlaw)}, with an extra-pruning of the marginal summary statistics at each round and retraining from scratch.}
    \label{fig:pruning-training}
\end{figure}

\newpage

\section{Application to true spectra}

\begin{figure*}[h!]
    \centering
    \includegraphics[width=0.49\linewidth]{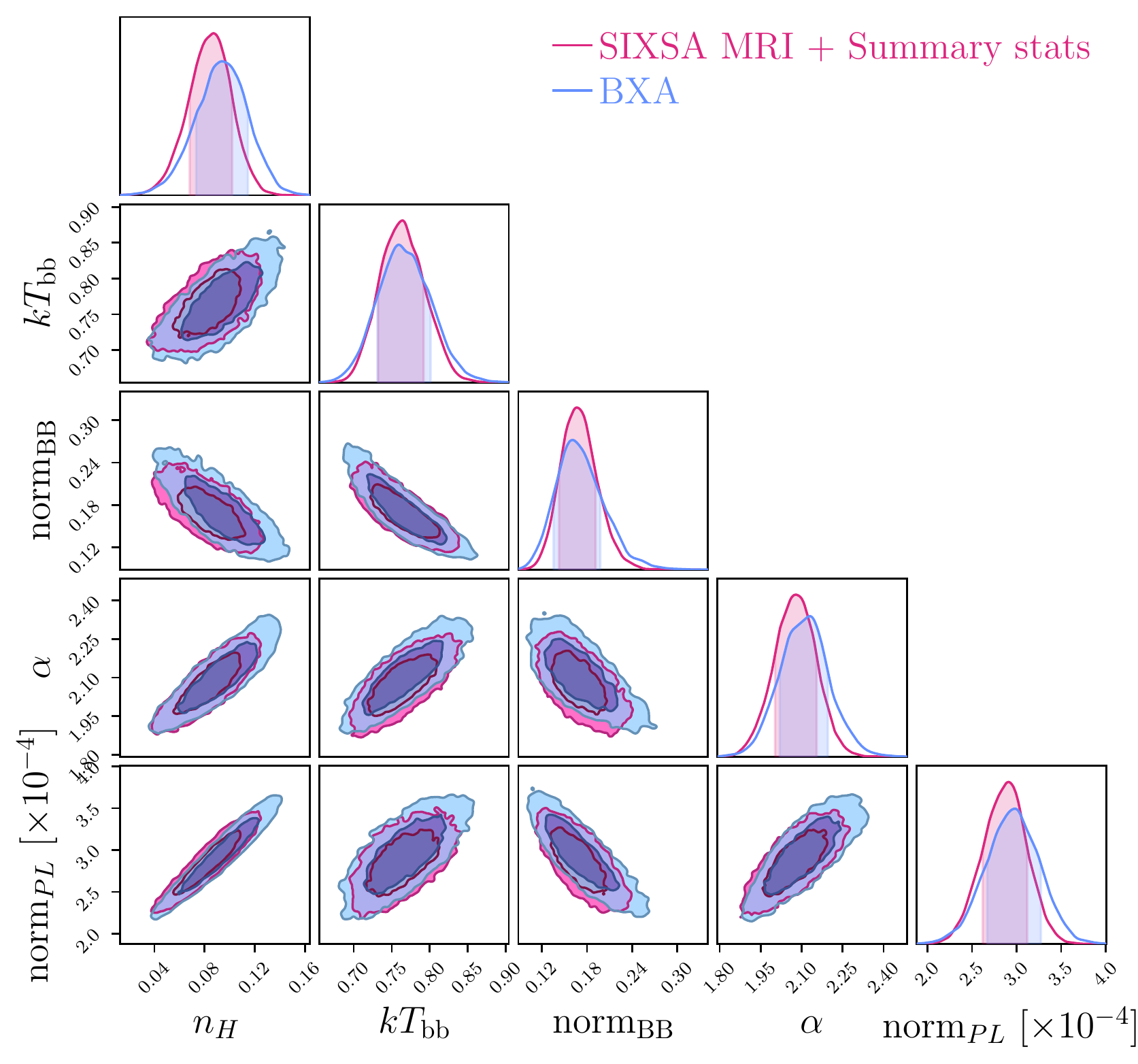}
    \includegraphics[width=0.49\linewidth]{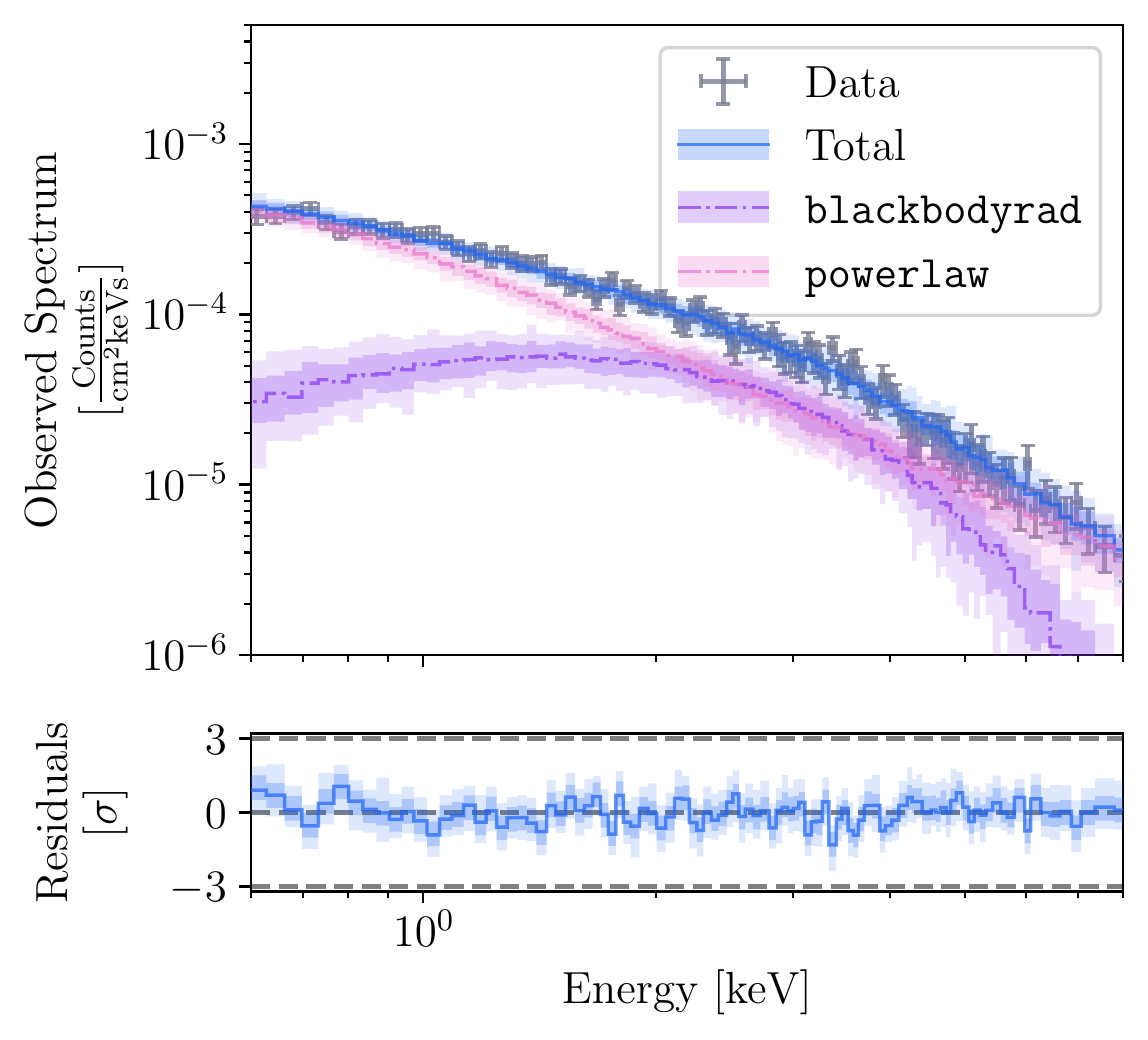}
    \caption{Left: Comparison of the posterior distributions obtained from a \texttt{BXA} run with 1k live-points and MRI run using summary stats for the spectrum of NGC 7793 ULX-4. Right: Posterior predictive spectra associated with the posterior distribution obtained using summary statistics. The bands detail the (16-84) and (2.5-97.5) percentiles of the posterior predictive spectra.}
    \label{fig:true-data}
\end{figure*}

To assess the applicability of summary statistics to real observational data, we apply our methodology to the same ULX spectrum used as a reference in \cite{Dupourque2024A&A...690A.317D}. This dataset corresponds to an \textit{XMM-Newton} observation of NGC 7793 ULX-4, previously reduced and analysed by \cite{2021MNRAS.503.5485Q}, and features a high-quality EPIC-PN spectrum with no background subtraction. The source emission is well described by an absorbed dual-component model (\texttt{tbabs*(powerlaw+blackbodyrad)}), typical for ULXs exhibiting both thermal and non-thermal components. Using the same inference setup as in the previous work, we apply both \texttt{BXA} and a MRI approach with 5 rounds of 5,000 simulations each. The posterior predictive and comparison of posterior distributions are shown in Fig.~\ref{fig:true-data}, further validating the robustness of summary statistics on real astrophysical data.

\newpage

\section{Extra figures}

\begin{figure*}[h!]
    \centering
    \includegraphics[width=0.7\linewidth]{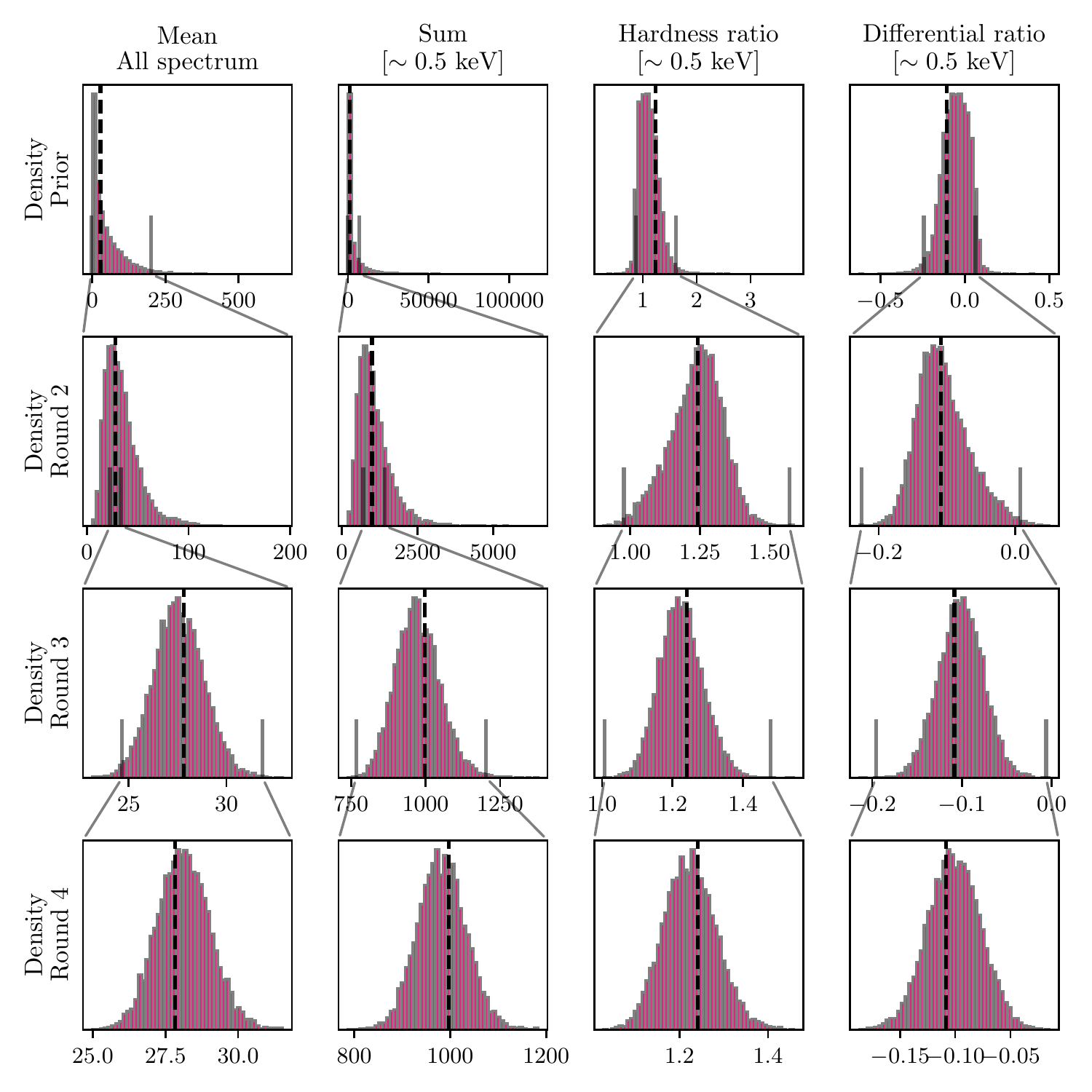}
    \caption{Evolution of the mean value of the spectrum, the sum, hardness ratio and differential ratios around 0.5 keV in the \texttt{tbabs*(comptt+powerlaw)} model. The true value is displayed in a black vertical line.}
    \label{fig:summary-rounds}
\end{figure*}

\begin{figure*}[t]
    \centering
    \includegraphics[width=0.7\linewidth]{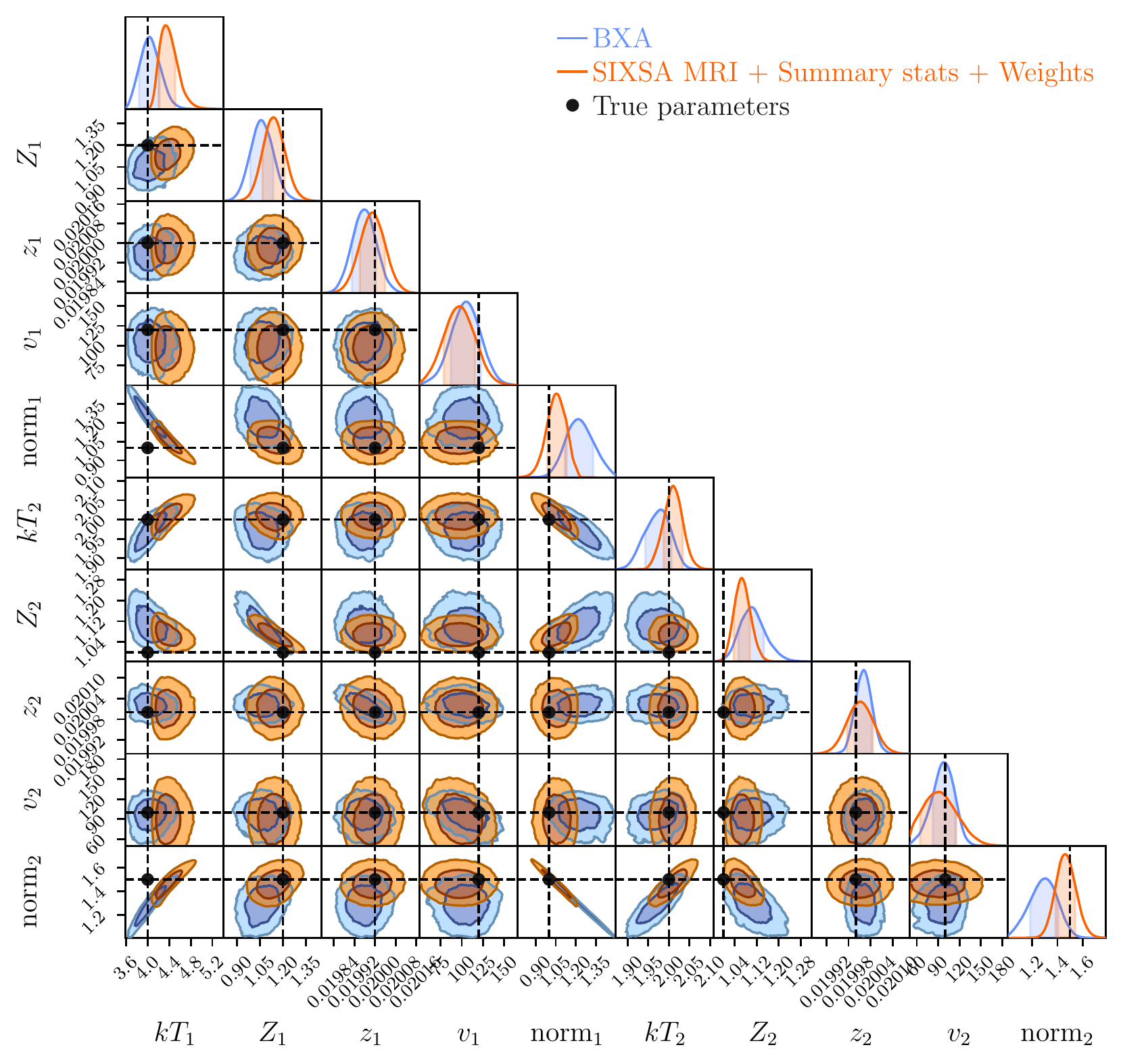}
    \vspace{0.2cm} 
    \includegraphics[width=0.7\linewidth]{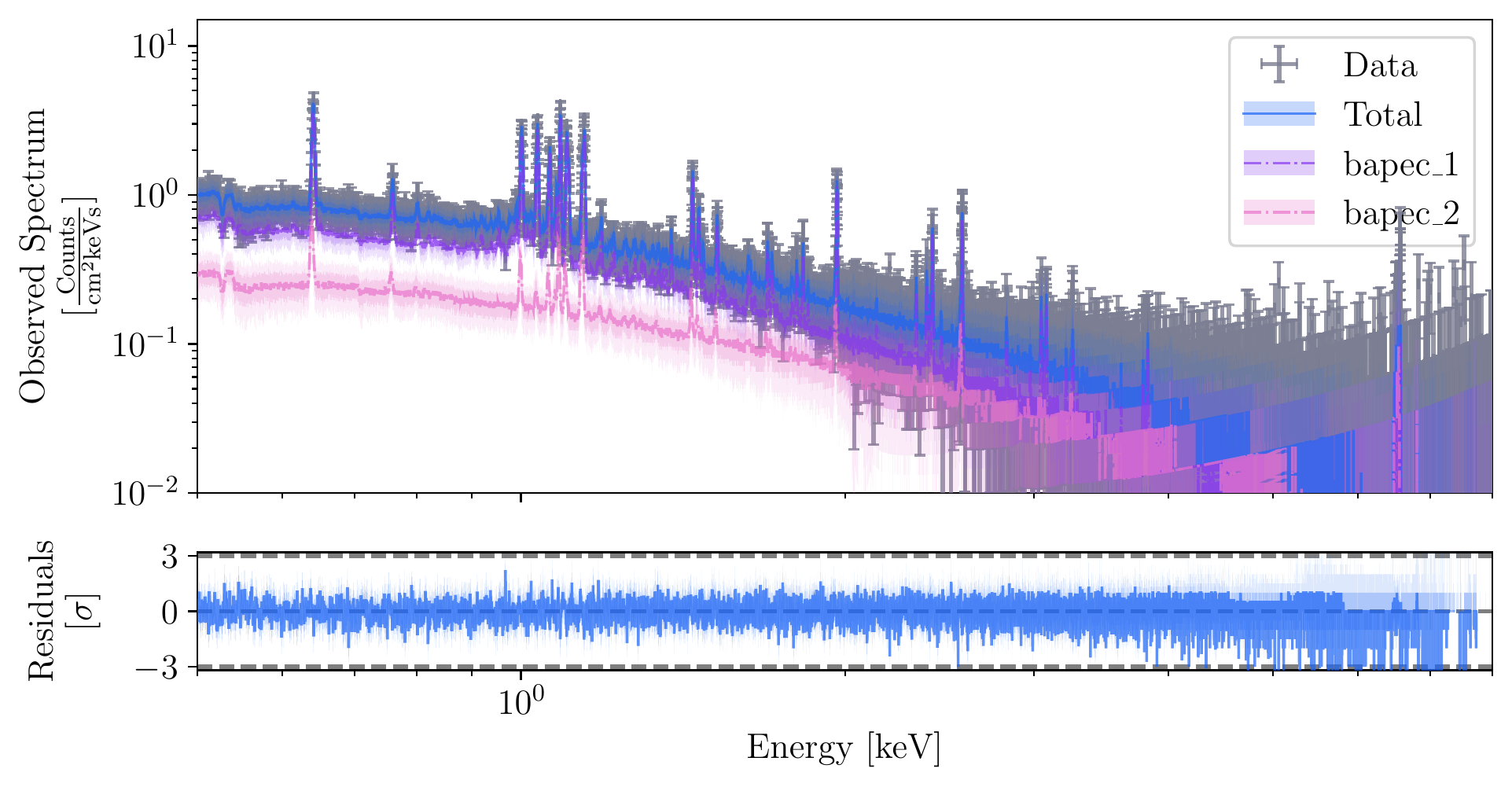}
    \caption{Top: Comparison of the posterior distributions obtained from a \texttt{BXA} run with 1k live-points and MRI run using summary stats \texttt{tbabs*(bapec+bapec)} model for 20 rounds of 10k simulations. Bottom: Posterior predictive spectra associated with the posterior distribution obtained using summary statistics and focus on 4 line complexes. The bands detail the (16-84) and (2.5-97.5) percentiles of the posterior predictive spectra.}
    \label{fig:apec-comparison}
\end{figure*}

\begin{figure*}[t]
    \centering
    \includegraphics[width=\linewidth]{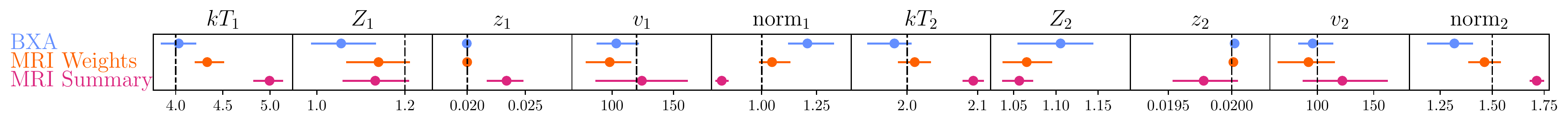}
    \caption{Comparison of the posterior marginal distributions obtained from a \texttt{BXA} run with 1k live-points, MRI run using summary stats and extra weighted energies from complexes \texttt{tbabs*(bapec+bapec)} model for 20 rounds of 10k simulations and MRI in the same setup with only summary stats.}
    \label{fig:apec-full-comparison}
\end{figure*}
\FloatBarrier 
\twocolumn
\end{document}